\documentclass[a4paper,fleqn,usenatbib]{mnras}

\usepackage{mathptmx}

\usepackage[T1]{fontenc}
\usepackage{ae,aecompl}

\usepackage{graphicx}	
\usepackage{amsmath}	
\usepackage{amssymb}	
\usepackage{natbib,epsfig,graphicx,ulem}
\usepackage{txfonts}
\usepackage{longtable}
\usepackage{soul,color,lscape}

\usepackage{etoolbox}
\makeatletter
\patchcmd\@combinedblfloats{\box\@outputbox}{\unvbox\@outputbox}{}{%
   \errmessage{\noexpand\@combinedblfloats could not be patched}%
}%
 \makeatother

\newcommand{\nic}[1]{#1}

\title[KiDS-450: Cosmological Constraints from Weak Lensing Peak Statistics - II]{KiDS-450: Cosmological Constraints from Weak Lensing Peak Statistics - II: Inference from Shear Peaks using N-body Simulations}

\author[Martinet et al.]{Nicolas Martinet$^{1}$\thanks{E-mail: nmartinet@astro.uni-bonn.de},
Peter Schneider$^{1}$,
Hendrik Hildebrandt$^{1}$,
HuanYuan Shan$^{1}$,\newauthor
Marika Asgari$^{2}$,
J\"org P. Dietrich$^{3,4}$,
Joachim Harnois-D\'eraps$^{2}$,
Thomas Erben$^{1}$,\newauthor
Aniello Grado$^{5}$,
Catherine Heymans$^{2}$,
Henk Hoekstra$^{6}$,
Dominik Klaes$^{1}$,\newauthor
Konrad Kuijken$^{6}$,
Julian Merten$^{7,8}$,
Reiko Nakajima$^{1}$
\\
$^{1}$Argelander-Institut f\"ur Astronomie, Universit\"at Bonn, Auf dem H\"ugel 71, D-53121 Bonn, Germany\\
$^{2}$Institute for Astronomy, University of Edinburgh, Royal Observatory, Blackford Hill, Edinburgh EH9 3HJ, UK\\
$^{3}$Faculty  of  Physics,  Ludwig-Maximilians-Universit\"at,  Scheinerstr.  1, 81679 Munich, Germany\\
$^{4}$Excellence  Cluster  Universe,  Boltzmannstr.  2,  85748  Garching,  Germany\\
$^{5}$Astronomico di Capodimonte, Via Moiariello 16 80131 Napoli Italy\\
$^{6}$Leiden Observatory, Leiden University, P.O.Box 9513, 2300RA Leiden, The Netherlands\\
$^{7}$Department of Physics, University of Oxford, Keble Road, Oxford OX1 3RH, UK\\
$^{8}$INAF, Osservatorio Astronomico di Bologna, via Pietro Gobetti 93/3, 40129 Bologna, Italy}

\date{Accepted XXX. Received YYY; in original form ZZZ}

\pubyear{2017}

\begin{document}
\label{firstpage}
\pagerange{\pageref{firstpage}--\pageref{lastpage}}
\maketitle

\begin{abstract}

\noindent We study the statistics of peaks in a weak lensing reconstructed mass map of the first 450 square degrees of the Kilo Degree Survey. The map is computed with aperture masses directly applied to the shear field with an NFW-like compensated filter. We compare the peak statistics in the observations with that of simulations for various cosmologies to constrain the cosmological parameter $S_8 = \sigma_8 \sqrt{\Omega_{\rm m}/0.3}$, which probes the ($\Omega_{\rm m}, \sigma_8$) plane perpendicularly to its main degeneracy. We estimate $S_8=0.750\pm0.059$, using peaks in the signal-to-noise range $0 \leq {\rm S/N} \leq 4$, and accounting for various systematics, such as multiplicative shear bias, mean redshift bias, baryon feedback, intrinsic alignment, and shear-position coupling. These constraints are $\sim25\%$ tighter than the constraints from the high significance peaks alone ($3 \leq {\rm S/N} \leq 4$) which typically trace single-massive halos. This demonstrates the gain of information from low-S/N peaks. However we find that including ${\rm S/N} < 0$ peaks does not add further information. Our results are in good agreement with the tomographic shear two-point correlation function measurement in KiDS-450. Combining shear peaks with non-tomographic measurements of the shear two-point correlation functions yields a $\sim20$\% improvement in the uncertainty on $S_8$ compared to the shear two-point correlation functions alone, highlighting the great potential of peaks as a cosmological probe.
\end{abstract}



\defcitealias{Shan+17}{Paper I}

\section{INTRODUCTION}
\label{sec:intro}

In a recent study, \citet{Hildebrandt+17} measured the coherent lensing distortions of galaxy images by large-scale structures (LSS) as a function of angular separation in the first 450 square degrees of the Kilo Degree Survey (hereafter KiDS-450). This cosmic shear analysis yielded an $S_8 ~(=\sigma_8\sqrt{\Omega_{\rm m}/0.3})$ value that is 2.3$\sigma$ lower than that inferred from Planck Cosmic Microwave Background (CMB) measurements \citep{PlanckXIII}. This difference between low- and high-redshift probes, if it is not due to systematic effects or a statistical fluctuation, may point to new physics. To improve the constraints, we propose to use the statistics of peaks in the weak lensing (WL) mass map of KiDS-450 in order to infer an additional lensing measurement of $S_8$, based on a different statistic than shear two-point correlation functions (hereafter 2PCFs).

The distribution of peak heights in mass maps depends on cosmology. In particular, peaks are sensitive to the matter density $\Omega_{\rm m}$ and the amplitude of the matter power spectrum described by $\sigma_8$ on scales of 8~$h^{-1}$Mpc, as these parameters impact the mass and the abundance of Dark Matter (DM) halos. Peak statistics has been successfully used either to predict achievable cosmological constraints \citep[e.g. ][]{DH10,Kratochvil+10,Yang+11,Hilbert+12,Marian+12,Marian+13,Martinet+15b} or to directly measure them from observations \citep[e.g. ][]{Liu+15j,Liu+15x,Kacprzak+16}. 

In contrast to classical 2nd-order cosmic shear probes, shear peaks are sensitive to the non-Gaussianities in the matter and shear distributions. Commonly, while large peaks correspond to single massive halos, the lower-amplitude peaks are often due to the projection of multiple smaller halos \citep{Yang+11,LiuJ+16} and are also sensitive to cosmology \citep{Jain+00,Wang+09,DH10,Kratochvil+10}. Low-amplitude peaks can also be produced by mass outside collapsed DM halos or by shape noise contribution, and it is not clear yet which of these three different origins is dominant. Although 2nd-order cosmic shear and peak statistics do not probe the exact same information, they are both sensitive to LSS, and their cosmological constraints are correlated \citep[e.g ][]{DH10,Liu+15j}. As these two methods use the same observables but different statistics, comparing their respective constraints is a good test for method-dependent systematics.

Peak statistics have been analysed with various methods. The main differences between studies arise from both measurement and modelling choices. From the measurement point of view one can choose to reconstruct the WL map in convergence \citep[e.g. ][]{Kratochvil+10,Yang+11,Shan+14,Liu+15j,Petri+16} or shear space through compensated filters \citep[e.g. ][]{Kruse+99,Kruse+00,DH10,Maturi+11,Hamana+12,Martinet+15b,Kacprzak+16}. The shear approach properly deals with the mass sheet degeneracy, which is only approximately handled in the convergence case. See \citet{Lin+17} for a recent comparison of the cosmological parameter estimates from peaks computed in shear and convergence spaces. Furthermore, working in shear space allows one to include the observational masks, at the cost of computational time, as it requires to drop the Fourier Transform approach.

The modelling of the peak distribution can be done with either simulations or analytical predictions. N-body simulations capture the non-linear regime of structure formation allowing the use of the full signal-to-noise (S/N) range of peaks. Although most studies rely on simulations, analytical predictions based on the halo mass function offer a promising way to speed up peak studies \citep{Fan+10,Lin+15,Shirasaki17}. In particular, \citet{Zorrilla+16} showed that the model from \citet{Lin+15} predicts the mean abundance of high-S/N peaks reasonably well compared to N-body simulations, but that further development is needed to accurately estimate their variance or to probe the low-S/N tail. Both simulations and analytical predictions need to be adapted to the studied survey to capture the full complexity of the data.

In this paper we apply aperture masses \citep{Schneider96,BS01} in shear space. We compare the peak distribution from the KiDS-450 data to the \citet{DH10} simulations for various cosmologies and infer cosmological constraints on $S_8$. We also use mock data from the Scinet Light Cone Simulations \citep[SLICS: ][]{Harnois-Deraps+15} to refine our covariance matrix and estimate the impact of sample variance. Measuring the mass maps for various filter scales, we assess the gain of information from a multi-scale analysis. We compare our constraints on $S_8$ to the KiDS tomographic cosmic shear results \citep{Hildebrandt+17} in the context of the tension with Planck. Finally, we measure the non-tomographic shear 2PCFs and present joint constraints for peaks and 2PCFs.

This paper is the second in a series of papers on peak statistics in KiDS-450. \citeauthor{Shan+17} \citepalias[\citeyear{Shan+17}; hereafter][]{Shan+17} conducted an analogous analysis in convergence space, predicting the abundance of high-S/N peaks from an analytical model adapted from \citet{Fan+10}. The use of simulations allows us to additionally probe the information contained in the low-S/N peaks, at the cost of only sparsely sampling the ($\Omega_{\rm m}, \sigma_8$) cosmological plane. These two different approaches allow us to derive robust cosmological constraints from the peak statistics of the KiDS-450 survey and represent the largest observational WL peak statistic analyses to date.

The paper is structured as follows. We describe our observations and simulations in Sect.~\ref{sec:data} and Sect.~\ref{sec:sims} respectively. We then explain our mass map reconstruction in Sect.~\ref{sec:map} and present the KiDS peak distribution in Sect.~\ref{sec:peaks}. We estimate cosmological constraints in Sect.~\ref{sec:cc} and discuss them in Sect.~\ref{sec:discussion}.

\section{OBSERVATIONS}
\label{sec:data}

This analysis is based on the KiDS-450 data release, presented in \citet{Hildebrandt+17} and \citet{deJong+17}, and therefore uses the same input galaxy catalogue. The KiDS survey is also documented in \citet{deJong+15} and \citet{Kuijken+15} and we refer the reader to these papers for a complete description of the dataset and the reduction pipelines. Nevertheless, we summarise the main aspects of the survey and the steps in the reduction that are relevant for the present study.

KiDS is a ground-based survey optimised for WL measurements. The KiDS-450 sample is an intermediate release of the ongoing survey that covers 449.7 square degrees, split into five patches: three on the equatorial (G9, G12, and G15), and two in the southern sky (G23 and GS). Images are acquired with the OmegaCAM camera on the 2.6m VLT Survey Telescope, in four optical bands ({\it u},{\it g},{\it r}, and {\it i}). Weak lensing shape measurements are carried out on the {\it r}-band images which reach a limiting magnitude of 24.9 ($5\sigma$ in a 2 arcsec aperture) and have a median seeing of 0.66 arcsec. Galaxy shapes are determined with the updated version of the model fitting algorithm {\it lens}fit \citep{Miller+07}, described in \citet{FenechConti+17}. Photometric redshifts $z_{\rm B}$ are computed with the Bayesian code {\small BPZ} \citep{Benitez00} using the four optical bands and are described in \citet{Kuijken+15}. The redshift distribution is estimated from spectroscopically matched galaxies \citep{Hildebrandt+17}. We apply the same redshift cut as for the 2PCFs analysis: $0.1< z_{\rm B}\leq0.9$, but do not split the data into different redshift bins. This choice is driven by limitations on the simulation side, and is explained in Sect.~\ref{subsec:sims:adapt}. The total number of galaxies in our catalogue is $\sim12.34$ million after the redshift cuts.

For any shape measurement method one needs to calibrate the biases in the shear estimates. This is usually decomposed in a multiplicative and additive term in a linear relation between measured and true shear. The multiplicative bias of each galaxy is the same as in \citet{Hildebrandt+17}, and is estimated through extensive simulations in \citet{FenechConti+17}. As suggested in \citet{Miller+13}, it is better to correct for multiplicative bias in a global approach to avoid possible correlation between ellipticities and correction factors. This correction is described in Sect.~\ref{sec:map} and applied to each aperture mass in Eq.~(\ref{eq:SN}). We compute the mean additive shear bias as the average weighted ellipticity over all galaxies. The calculation is done independently for each of the five patches, and for each of the two ellipticity components. The values differ from those of \citet{Hildebrandt+17} because they determined it independently for several redshift slices while we use a single redshift bin. This bias is always lower than $1.5\times10^{-3}$, and is subtracted from the measured ellipticities.

\section{SIMULATIONS}
\label{sec:sims}

We derive cosmological constraints by comparing the WL peak distribution of KiDS-450 to that of simulations with varying cosmologies. To that purpose we use the simulations from \citet{DH10}. In Appendix~\ref{sec:discussion:cov} we also use mock catalogues from the SLICS simulations \citep{Harnois-Deraps+15} to better estimate the covariance matrix, and compare it with the covariance matrix from the \citet{DH10} simulations that is used for parameter inference.

\subsection{Dietrich \& Hartlap (2010) simulations}

The \citet{DH10} simulations consist of a set of 192 N-body simulations run with the {\small GADGET-2} software \citep{Springel05}, with initial conditions generated with the \citet{Eisenstein+98} transfer function. 256$^3$ dark matter particles are evolved from $z=50$ to $z=0$ in a box of $200 ~h_{70}^{-1}{\rm Mpc}$ side length, with particle mass varying between $9.3\times10^9{\rm M_\odot}\leq m_{\rm p} \leq 8.2\times10^{10}{\rm M_\odot}$, depending on the cosmology. Each simulation spans a $6 \times 6$~deg$^2$ field-of-view.

\begin{figure}
\centering
\includegraphics[width=0.7\textwidth,clip,angle=0]{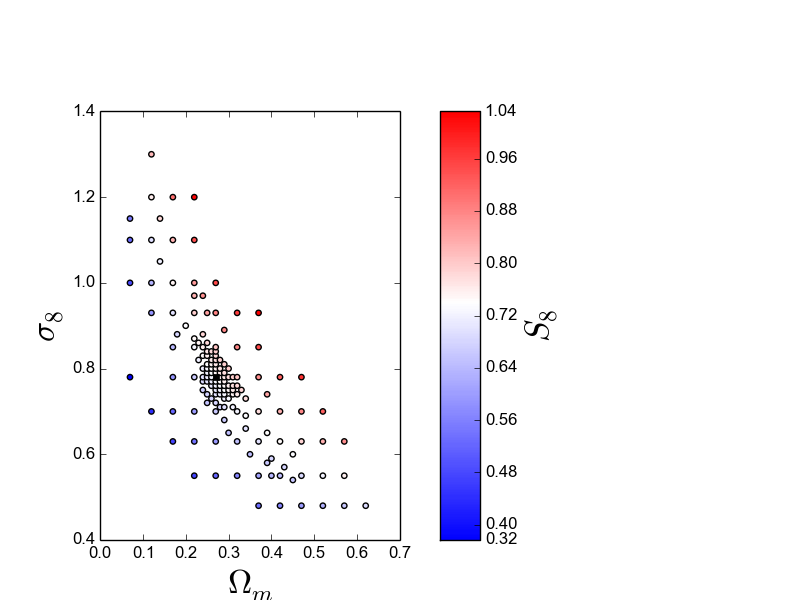}
\caption{Sampling of the ($\Omega_{\rm m}, \sigma_8$) plane. Each dot represents an N-body simulation out of which 5 galaxy catalogues are made. Colours show $S_8=\sigma_8\sqrt{\Omega_{\rm m}/0.3}$ values. The central black dot corresponds to our fiducial cosmology $\bmath{\pi_0}=(0.27, 0.78)$, which has 35 N-body simulations and therefore 175 pseudo-independent catalogues.}
\label{fig:cosmotile}
\end{figure}

These simulations are run with cosmological parameters $\bmath{\pi}=(\Omega_{\rm m}, \sigma_8)$. Among them, 35 are run with fiducial cosmological parameters $\bmath{\pi_0}=(0.27, 0.78)$, and 158 have $\Omega_{\rm m}$ and $\sigma_8$ spanning a large range of values. One set of simulations was lost due to an archiving issue, and we therefore only use 157 different cosmologies. As seen in Fig.~\ref{fig:cosmotile}, the steps in the ($\Omega_{\rm m}$, $\sigma_8$) plane are smaller around the fiducial parameters, allowing a better precision on the variation of the WL peak distribution around the cosmological parameter values expected from previous cosmological studies. We also show the variation of $S_8=\sigma_8\sqrt{\Omega_{\rm m}/0.3}$, which is the parameter to which we are most sensitive, given the degeneracy between $\Omega_{\rm m}$ and $\sigma_8$. The other parameters that are not probed in this study are fixed to their fiducial values ($\Omega_{\rm b}=0.04$, $n_{\rm s}=1.0$, and $h_{70}=1$), except $\Omega_{\rm \Lambda}$ which varies with $\Omega_{\rm m}$ to preserve flatness.

Ray-tracing is then performed through each simulation to produce convergence and shear maps, from which a catalogue of galaxies is generated through random position sampling. Random shifting within a simulation snapshot was used to extract 5 pseudo-independent ray-tracings out of a single N-body run. These mock catalogues mimic the Canada-France-Hawaii Telescope Legacy Survey (CFHTLS) in terms of redshift distribution, galaxy number density and shape noise.

Further details on the \citet{DH10} simulations and the creation of the mock catalogues can be found in the corresponding paper.

\subsection{Adapting to the KiDS survey}
\label{subsec:sims:adapt}

Because the \citet{DH10} simulations are not tailored for KiDS-450 data we need to modify their output. In particular we want to use the same positions, redshift distribution, and shape noise as in the data.

\begin{figure}
\centering
\includegraphics[width=0.5\textwidth,clip,angle=0]{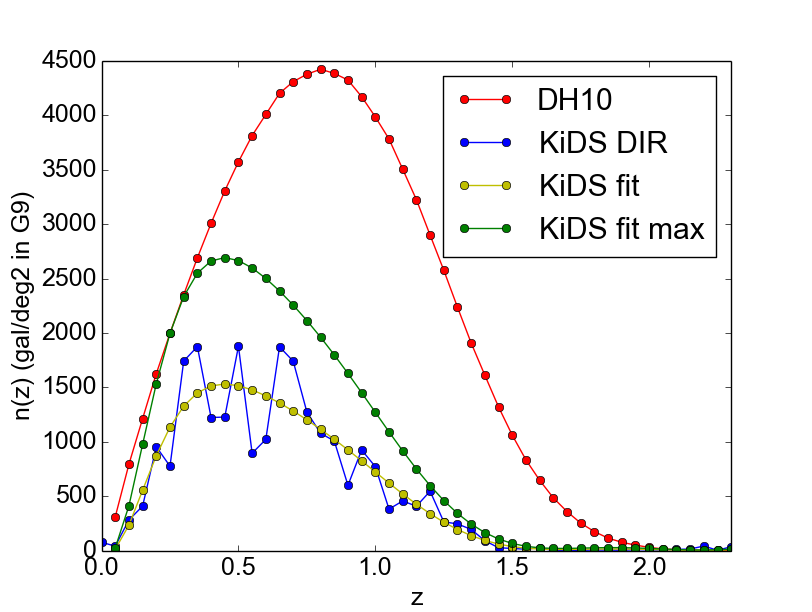}
\caption{Illustration on patch G9 of the down-sampling of the \citet{DH10} simulations to match KiDS redshift distribution. Red corresponds to the initial \citet{DH10} $n(z)$, blue to the KiDS DIR $n(z)$, yellow to the smoothed KiDS DIR $n(z)$, and green to the down-sampled \citet{DH10} redshift distribution that matches that of KiDS.}
\label{fig:adaptz}
\end{figure}

The first step is to modify the redshift distribution of the simulations by sub-sampling the galaxies in order to match the KiDS redshift distribution. This is possible because the mocks have a much higher galaxy density than KiDS, i.e. 25 versus $\sim 8.5$ galaxies per square arcminute. We use the DIR redshift distribution detailed in \citet{Hildebrandt+17} which corresponds to the redshift distribution of a magnitude-reweighed sample of spectroscopically-matched galaxies in the photometric redshift range $0.1<z_{\rm B}\leq0.9$. It was shown that this approach is more precise than using photometric redshifts, and this redshift distribution extends by construction above $z_{\rm spec}=0.9$. The KiDS galaxy density after applying this redshift cut is $\sim7.5$ galaxies per square arcminute. The process is illustrated in Fig.~\ref{fig:adaptz}. We first fit the KiDS DIR redshift distribution with a polynomial of 12th order chosen to smooth the distribution. We check that this fit does not change the mean redshift of the distribution. However, the \citet{DH10} mocks contain very few galaxies at $z>2$ due to the redshift distribution they adopted. Thus, we reject most galaxies selected in $0.1 < z_{\rm B} \leq 0.9$ with $z_{\rm spec} \geq 2$. This shifts the mean redshift by $\sim 0.05$ towards a lower value. We then look for the largest multiplicative factor that can be applied to this smoothed distribution in order not to exceed the distribution in the simulation at any $z$. Taking the ratio of this last distribution (the green points in Fig.~\ref{fig:adaptz}) to the $n(z)$ of the simulations (red points of Fig.~\ref{fig:adaptz}) gives a weight between 0 and 1 to each redshift bin. We finally down-sample the simulation drawing for each galaxy a random number between 0 and 1 and discarding the galaxy if this number is above the weight of the galaxy redshift bin.

We then use a nearest-neighbour approach to assign a simulated reduced shear value at each of the observed positions. One could also use a linear interpolation of the four simulated galaxies closest to the observed one that we try to match. This technique would be more accurate if the simulated galaxies were placed on a grid. However, these galaxies are at random positions, and could lead to combination of shears from source galaxies that are not affected by the same lenses. We therefore assign the shear of the closest neighbour. For each simulation we then build a catalogue of galaxies whose positions, weights, and intrinsic ellipticities are taken from the observed KiDS catalogue, and shears from the simulation. The KiDS-450 observational masks are also applied when assigning positions. The observed ellipticities are rotated by a random angle before being assigned to our simulated catalogue, allowing us to remove the signal from the observation but retaining its exact shape noise. The shear from the data slightly modifies the amplitude of the intrinsic ellipticity used in the simulations but this effect is small as the shear amplitude is of the order of a few percents of the ellipticity. We also bias simulated values of the shear by the multiplicative bias measured for the corresponding observed galaxy, so that the bias is consistent between observations and simulations. However, ignoring this bias in the simulation affects the final cosmological constraints by less than $0.01\sigma$. Since the peak distribution is very sensitive to shape noise, we make several noise realisations by applying different random rotations to the observed ellipticities. This point is discussed in more detail in Sect.~\ref{subsec:meas}.

There are three caveats to this interpolation scheme. The first is that the KiDS data cover 450 square degrees while each simulation is only $6\times6$ square degrees. We therefore have to use the same simulation several times to cover the entire observational field, which underestimates the sample variance. The effect of this procedure is studied in Appendix~\ref{sec:discussion:cov}, making use of the larger SLICS simulations \citep{Harnois-Deraps+15}. The second issue is that the galaxy density of the simulated mock catalogue is not large enough compared to that of the observations to ensure that no simulated shear value is used more than once in the interpolation process. As a consequence, some close galaxies in the matched catalogue will have the same shear (but different intrinsic ellipticities). However, this effect is mitigated by the fact that the separation between closest neighbours is much smaller than the scale of the filter that we are applying in the aperture mass calculation. Quantitatively, the mean separation between closest neighbours is $0.145$ arcmin with a standard deviation of $\pm0.076$ arcmin and the filter's outer and effective radii are 12.5 and 1.875 arcminutes, respectively. Even if a galaxy is attributed a shear from a slightly different position, this difference is not significant as seen by the filter function, leading to the same result as if the shear was estimated at the true galaxy position. This problem would become significant only if we were conducting a tomographic analysis, because the distance to the closest neighbour would become too large. A tomographic approach would thus require to directly build the mock catalogue at the desired positions and redshifts through looking up the values in the shear planes calculated at various redshifts in the simulations. Finally, the downsampling diminishes the correlations between the lensing mass distribution and the source galaxy distribution, as the observed redshifts are randomly re-shuffled. This effect is however accounted for as a systematic bias when evaluating the difference of S/N between simulated and observed peaks (Sect.~\ref{subsec:sys}).

The final simulation products consist of 175 catalogues at the fiducial cosmology and 785 catalogues at 157 different cosmologies. These catalogues have their shear values estimated from the \citet{DH10} simulations, and their positions, weights, and intrinsic ellipticities from the observations. We note that the simulations do not include the full complexity of the observations. In particular baryon feedback is not captured by these DM only simulations and the lens-source coupling is lost when assigning observed galaxy positions to the mocks. The impact of these effects on cosmological constraints is discussed in terms of systematics in Sect.~\ref{subsec:sys}.

\subsection{KiDS SLICS mocks}
\label{subsec:slics}

In Appendix~\ref{sec:discussion:cov}, we use the SLICS simulations to refine the covariance matrix and study the impact of sample variance on the cosmological constraints. In the rest of the paper, the mocks built from the \citet{DH10} simulations are used. The SLICS simulations \citep{Harnois-Deraps+15} consist of 930 N-body simulations with $1536^3$ particles evolved in a box of 505~$h^{-1}$Mpc, and cover $10 \times 10$ square degrees in the redshift range $0<z<3$. Each particle has a mass of $2.88\times10^9 {\rm M_\odot} h^{-1}$. Every simulation has the same cosmology: $\Omega_{\rm m}=0.2905, \Omega_{\rm \Lambda}=0.7095, \Omega_{\rm b}=0.0473$, $h=0.6898$, $\sigma_8=0.826$, and $n_{\rm s}=0.969$, but different initial conditions.

As described in \citet{Hildebrandt+17}, mock galaxy catalogues are drawn from these simulations, estimating the shear at various positions over 18 redshift planes. In addition to several improvements of the simulation quality compared to the \citet{DH10} simulations, these mocks estimate the shear at the observed galaxy position without resorting to interpolation. This is also in contrast with the mocks used in \citet{Hildebrandt+17} where galaxies are at random positions. We have verified from the \citet{DH10} simulations that using shear instead of reduced shear does not significantly affect the cosmological constraints derived from our peak estimator. We therefore use shear instead of reduced shear from the SLICS simulations,  making the calculation faster.

From this set of simulations we make 67 independent realisations of the KiDS-450 footprint, using different simulations to tile the space. This means that in contrast to the mocks we build from the \citet{DH10} simulations which map the full 450~deg$^2$ of data with 36~deg$^2$ of simulations, these refined mocks better account for sample variance, as 450~deg$^2$ of simulations are used to map the 450~deg$^2$ of data. Details on the tiling will be available in a forthcoming paper (Harnois-D\'eraps et al. 2017, in prep.).

\section{APERTURE MASS CALCULATION}
\label{sec:map}

Peaks are detected in a map of aperture masses \citep{Schneider96,BS01}. This technique presents several advantages over the classical mass reconstruction from shear. In particular, it avoids the integration over finite area which introduces an unknown constant, due to the so-called mass sheet degeneracy. As the mass sheet degeneracy affects the signal but not the noise, its main effect is to add a random shift, different from field to field, to the S/N distribution of peaks. It also allows one to analytically compute local noise and to deal with masks in a simple fashion. This led to its extensive use in WL peak analyses \citep[e.g. ][]{DH10,Marian+12,Martinet+15b,Kacprzak+16}. In \citetalias{Shan+17}, the mass map is reconstructed through a shear-convergence inversion \citep{KS93} because it is simpler to model the analytical prediction of peaks in convergence space. However, it is preferable to use the aperture mass statistics as we do in this second paper, to avoid mass sheet degeneracy, and to better handle the masks.

The aperture mass is an integral of the local mass density around position $\bmath{\theta}_0$, weighted by a filter function which is compensated in the convergence space:

\begin{equation}
M_{\rm ap}(\bmath{\theta}_0) = \int {\rm d}^2 \bmath{\theta} \;U(\bmath{\theta}-\bmath{\theta}_0)\;\kappa(\bmath{\theta}),
\end{equation}

\noindent where the compensation of the isotropic weight function $U(\bmath{\theta})$ is expressed as: 

\begin{equation}
\int {\rm d} \theta \; \theta \; U(\theta) = 0.
\end{equation}

\noindent This condition ensures the aperture mass is insensitive to the (linear version of the) mass sheet degeneracy. For any compensated filter in convergence space $U(\bmath{\theta})$, one can compute the equivalent filter $Q(\bmath{\theta})$ in shear space, which gives the aperture mass from the tangential shear \citep{Schneider96}:

\begin{equation}
Q(\theta)=\frac{2}{\theta^2} \int_{0}^{\theta} {\rm d} \theta' \theta' U(\theta') - U(\theta),
\end{equation}

\begin{equation}
M_{\rm ap}(\bmath{\theta}_0) = \int {\rm d}^2 \bmath{\theta} \;Q(\bmath{\theta}-\bmath{\theta}_0) \; \gamma_{\rm t}(\bmath{\theta},\bmath{\theta}_0),
\end{equation}

\noindent where the tangential shear $\gamma_{\rm t}(\bmath{\theta},\bmath{\theta}_0)$ is expressed as a function of both shear components and the angle between the position where the shear is measured and the centre of the aperture $\phi(\bmath{\theta},\bmath{\theta}_0)$:

\begin{equation}
\gamma_{\rm t}(\bmath{\theta},\bmath{\theta}_0)= - \left[ \; \gamma_1(\bmath{\theta}) \cos(2\phi(\bmath{\theta},\bmath{\theta}_0)) + \gamma_2(\bmath{\theta}) \sin(2\phi(\bmath{\theta},\bmath{\theta}_0)) \; \right].
\end{equation}

In order to apply aperture masses to observed data, the integration is transformed into a sum over discrete positions where the shear is estimated, i.e. at galaxy positions $\bmath{\theta_i}$. The tangential shear is also replaced by the galaxy tangential ellipticity:

\begin{equation}
\label{eq:map}
M_{\rm ap}(\bmath{\theta}_0)=\frac{1}{n_{\rm gal}}\sum_i Q(\bmath{\theta}_i-\bmath{\theta}_0)\epsilon_{{\rm t}}(\bmath{\theta}_i, \bmath{\theta}_0),
\end{equation}

\noindent where $n_{\rm gal}$ is the galaxy density inside the aperture. The masks are easily handled as long as the computation is done in real space, as the masked galaxies can simply be ignored in the computation. However it significantly increases the computational time compared to Fourier space. We prioritise the exact handling of masks and therefore do the calculation in real space.

Galaxy ellipticity is equal to the reduced shear on average, provided that source galaxies are randomly oriented. This property is the fundamental hypothesis of WL and allows us to replace shear by ellipticity in Eq.~(\ref{eq:map}), also enabling the analytic computation of the local noise as the standard deviation of the aperture mass in the absence of shear:

\begin{equation}
\label{eq:N}
\sigma(M_{\rm ap}(\bmath{\theta}_0))=\frac{1}{\sqrt{2}n_{\rm gal}}\left(\sum_{i}{|\epsilon(\bmath{\theta}_i)|^{2}Q^{2}(\bmath{\theta}_i-\bmath{\theta}_0)}\right)^{1/2}.
\end{equation}

\noindent The sum over the squared ellipticity norm $\left( |\epsilon(\bmath{\theta}_i)|=\sqrt{\epsilon_1(\bmath{\theta}_i)^2+\epsilon_2(\bmath{\theta}_i)^2}\right)$ is sometimes replaced by the two dimensional dispersion of the ellipticity over the whole survey, and denoted by $\sigma_{\epsilon}$. However, it is more accurate to compute the shape noise at the level of each aperture as it varies from field to field, either for instrumental or physical reasons, e.g. varying depth, PSF variations or intrinsic alignments. We define the signal-to-noise (S/N) of each aperture as the ratio of $M_{\rm ap}$ and $\sigma(M_{\rm ap})$. Taking {\it lens}fit shear weights $w$ into account we can write this S/N as:

\begin{equation}
\label{eq:SN}
\frac{S}{N} \left( \bmath{\theta}_0 \right) = \frac{\sqrt{2} \sum_i Q(\bmath{\theta}_i-\bmath{\theta}_0) w(\bmath{\theta}_i) \epsilon_{{\rm t}}(\bmath{\theta}_i, \bmath{\theta}_0)}{\sqrt{\sum_{i}{w(\bmath{\theta}_i)^{2}|\epsilon(\bmath{\theta}_i)|^{2}Q^{2}(\bmath{\theta}_i-\bmath{\theta}_0)}}} \frac{\sum_i w(\bmath{\theta}_i)}{\sum_i w(\bmath{\theta}_i) \left[ 1+m(\bmath{\theta}_i)\right]}.
\end{equation}

As already stated in Sect.~\ref{sec:data}, the shear multiplicative bias correction is applied as the average weighted correction over every galaxy multiplicative bias $m(\bmath{\theta}_i)$ in the aperture. The correction appears as a normalisation to $M_{\rm ap}(\bmath{\theta}_0)$: $\sum_i w(\bmath{\theta}_i) \left[1+m(\bmath{\theta}_i)\right]$, but does not apply to $\sigma(M_{\rm ap}(\bmath{\theta}_0))$ which is only normalised by the sum over the galaxy weights: $\sum_i w(\bmath{\theta}_i)$. This is because the multiplicative bias is computed as a shear correction and the aperture mass noise is only sensitive to the intrinsic ellipticities.

As seen in the equations the aperture mass depends on the filter function $Q(\bmath{\theta})$. As we want to capture the signal from dark matter halos, we choose a shape that matches the expected tangential shear signal of a typical halo. While an NFW profile \citep{NFW97} would work well, we prefer to use an approximation of this profile to speed up the computation, namely the \citet{Schirmer+07} filter function:

\begin{equation}
\label{eq:Q}
\begin{aligned}
Q(\theta) =  & \left[1 + \exp \left(6 -150 \frac{\theta}{\theta_{\rm ap}}\right) + \exp \left(-47 +50 \frac{\theta}{\theta_{\rm ap}}\right)\right]^{-1} \\
             & \times \left(\frac{\theta}{x_{\rm c}\theta_{\rm ap}}\right)^{-1} \tanh \left(\frac{\theta}{x_{\rm c}\theta_{\rm ap}}\right),
\end{aligned}
\end{equation}

\noindent where $\theta_{\rm ap}$ is the radius of the aperture, and $x_{\rm c}$ is analogous to the halo concentration in the NFW profile, and is set to $x_{\rm c}=0.15$, found to be the optimal value for galaxy cluster detection \citep{Hetterscheidt+05}. The first term corresponds to an exponential cutoff at $\theta \longrightarrow 0$ and $\theta \longrightarrow \infty$. The cutoff at $\theta \longrightarrow 0$ is particularly important to avoid assigning too much weight to galaxies close to the aperture centre where reduced shear values may not be in the WL regime. The size of the filter is also important as it can preferentially select smaller or larger halos. In this study we set the fiducial aperture radius to $\theta_{\rm ap}=12.5$ arcmin, which maximises the number of peaks at ${\rm S/N} \geq 3$ in the KiDS data. With the chosen $x_{\rm c}$ parameter, this size corresponds to an effective radius $x_{\rm c}\theta_{\rm ap}$ of 1.875 arcmin. This aperture size gives the maximal sensitivity to massive halos. In Sect.~\ref{sec:discussion:multi}, we compute the peak distribution for different filter sizes and discuss correlations between scales and the potential gain of information from a multi-scale analysis.

We compute the aperture mass on a grid which covers the KiDS-450 area with a pixel size of 0.59 arcmin. This pixel size is a good trade-off between computational time and accuracy, as decreasing the pixel size further does not lead to the appearance of smaller structures. We discard all pixels closer to the edges of the reconstructed map than the aperture radius to avoid including incomplete apertures. However we note that these cuts do not significantly affect the cosmological parameter estimates as they are also applied to the simulations which have the same galaxy positions and masks. Maps are made independently for each patch: G9, G12, G15, G23, and GS \citep[see ][]{Hildebrandt+17}. Due to the incomplete current tiling of the survey we also subdivide G12, G15, and GS in 4, 3, and 2 sub-patches respectively, to avoid unnecessary computation in empty areas. 

\begin{figure}
\centering
\includegraphics[width=0.54\textwidth,clip,angle=0]{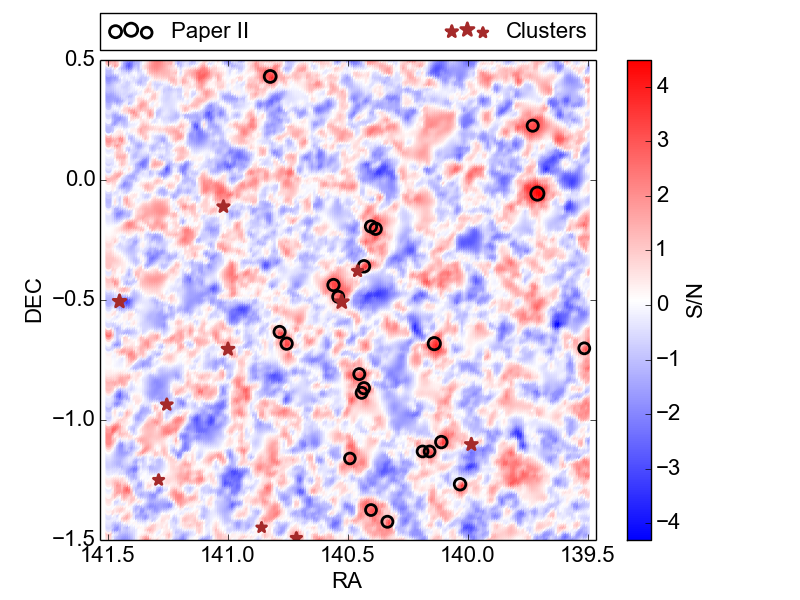}
\includegraphics[width=0.54\textwidth,clip,angle=0]{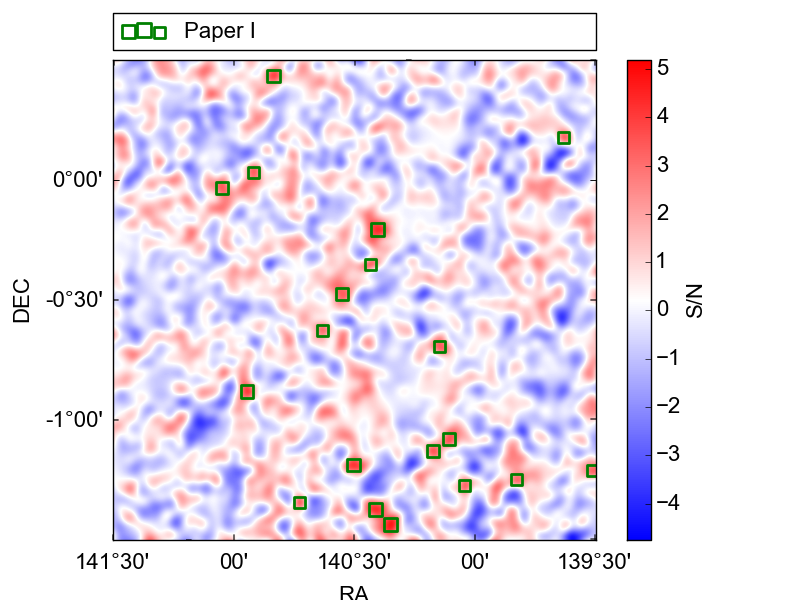}
\caption{{\it Top:} Aperture mass map of a $2\times2$ square degree field in the KiDS-450 footprint. The S/N of each pixel is colour-coded from blue to red, red corresponding to high-$M_{\rm ap}$ pixels. Black circles represent peaks in this map with ${\rm S/N} \geq 3$, and brown stars indicate galaxy clusters from \citet{Radovich+17} with redshift $z\leq0.5$ and detection level $\sigma \geq 7$. {\it Bottom:} Mass map of the same field computed from the direct shear inversion in \citetalias{Shan+17}. Green squares correspond to peaks with ${\rm S/N}\geq3$ as detected in \citetalias{Shan+17}.}
\label{fig:map}
\end{figure}

As an example of our reconstruction method, we show an aperture mass map for a $2\times2$ square degree field in the KiDS-450 footprint in Fig.~\ref{fig:map} and the detected peaks with ${\rm S/N} \geq 3$, defined as pixels with greater S/N than their 8 neighbours. For comparison, we also display the mass map and peaks with ${\rm S/N} \geq 3$ from \citetalias{Shan+17} for the same field. This second mass map is computed from a shear inversion method, with a single noise value across the survey. We see that the two maps trace the same LSS but present slight differences on small scales.  There is in particular a higher amount of substructures in the aperture mass map compared to the shear-inverted convergence map. This is probably due to the choice of the smoothing filter, which is an NFW-like $1.875\arcmin$ filter for aperture mass and a Gaussian $2\arcmin$ filter in the shear-inversion method. We also find that the peaks from each method do not all overlap due to the differences in the map computation and in the definition of the noise. Although the peak distributions from \citetalias{Shan+17} and the present study are different, the cosmological constraints should be comparable as the modelled peak distributions are computed in a consistent way with the observed distribution for each study.

We also compare our aperture mass map with known galaxy clusters overlapping with the KiDS-450 area \citep{Radovich+17}. These clusters have been detected through a matched filtering technique taking into account the magnitude distribution and density profile. We only retain clusters that are at redshift $z\leq0.5$ because higher-redshift clusters are unlikely to create a strong shear signal given the mean redshift of the background source population. We also cut out clusters that are detected with less than 7$\sigma$ significance to have a very pure sample. We see from Fig.~\ref{fig:map} that there is not a one-to-one correspondence between peaks and clusters. Only a few clusters are associated with peaks, but most clusters coincide with a high-S/N area of the WL mass map. This highlights that even at ${\rm S/N} \geq 3$ many peaks are not associated with clusters and contain a significant contribution from projection of low-mass halos or shape noise contamination. We also note that adding less significant clusters does not qualitatively change these conclusions. Finally, we recall that even if the aperture mass is computed with an NFW filter to match halos, our method is not optimised to cluster detection. In particular, we are sensitive to the integrated contribution along the line of sight which dilutes the signal from galaxy clusters.

\section{PEAK DISTRIBUTION}
\label{sec:peaks}

\subsection{Measurement}
\label{subsec:meas}

\begin{figure*}
\centering
\includegraphics[width=1.0\textwidth,clip,angle=0]{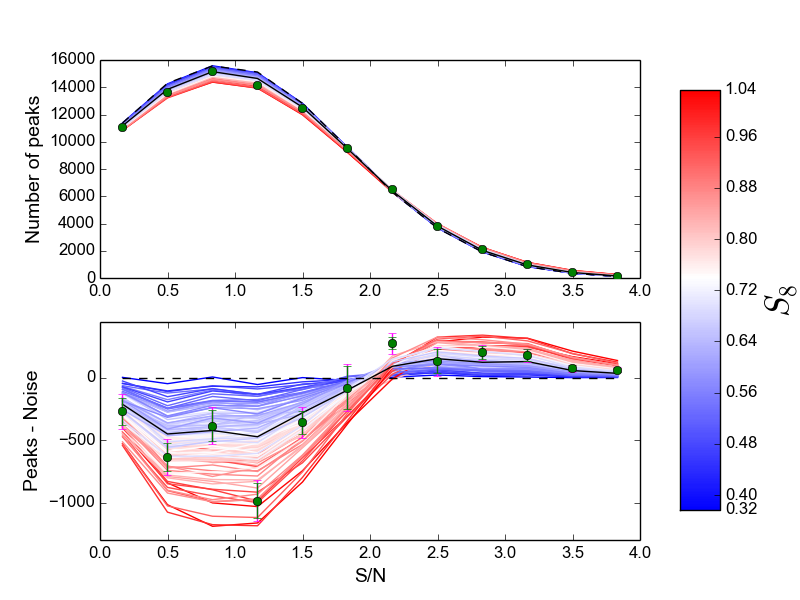}
\caption{Peak distribution ({\it top}) and differential peak distribution ({\it bottom}). The differential distribution corresponds to the peak distribution from which the averaged distribution over 5 noise-only field is subtracted. Green dots represent KiDS-450 data with error bars from the diagonal elements of the covariance matrix, the black line represents the mean of the fiducial cosmologies, black dashed line the noise only distribution and coloured lines the various simulations with $S_8$ increasing from blue to red. Error bars from bootstrap resampling of the data are displayed in magenta.}
\label{fig:distripeak}
\end{figure*}

Peaks are identified as pixels with S/N higher than their 8 neighbours in the aperture mass map, with the pixel scale of 0.59 arcmin. The global strategy is to measure the peak S/N distribution from the observations and the variation of the peak distribution with cosmology from the simulations.

Because we reproduce the same noise in the simulations as that of the observations, we can safely use any part of the peak distribution, including the low-S/N tail. However, the width of the S/N bins and the upper limit of the distribution must be chosen such to ensure that the distribution can be modelled by a multivariate Gaussian when computing the likelihood, i.e. that there is a sufficient number of peaks per bin. We note that this problem can also be dealt with by using the cumulative distribution \citep[e.g. ][]{DH10} or a varying width to get the same number of peaks per bin \citep[e.g. ][]{Martinet+15b}. However we use bins with fixed width because these other two methods would favour the more numerous low-S/N peaks given the chosen range of S/N. The number of bins is also limited by the precision we want to achieve on the covariance matrix. As shown in \citet{Taylor+14}, the more degrees of freedom the larger the uncertainty in the covariance. We use 12 bins of S/N equally spaced between 0 and 4, but also try a few other configurations (8 and 16 bins) to ensure that our constraints are insensitive to the bin width for reasonable choices. We refrain from adding peaks with ${\rm S/N} \geq 4$ as for these peaks the shear-position coupling becomes significant and can bias the results \citep{Kacprzak+16}. Shear-position coupling, also referred to as boost factor, biases the heights of peaks corresponding to large halos in the simulations compared to the observations because the redshift distribution of the data is applied to the simulations without prior knowledge of halo positions. This is described in more details in Sect.~\ref{subsec:sys} where we analyse the different systematic biases.

The error bars on the number of peaks displayed in the various figures correspond to the diagonal elements of the covariance matrix estimated from the fiducial cosmology mocks, based on the \citet{DH10} simulations. For the cosmological analysis the full covariance matrix is used. We also verify that these error bars are comparable to those computed through bootstrap resampling of the data. To estimate the bootstrap variances we divide the survey into 50 sub-patches with equal number of galaxies. Similarly to \citet{Hildebrandt+17} the definition of the sub-patches is based on right ascension cuts as the width in declination is roughly the same at any right ascension in the survey. This division leads to 50 patches which are roughly $3 \times 3$ square degrees. We then select 50 random patches, with the possibility of selecting the same patch more than once, to create a new peak distribution. Doing so 10,000 times and calculating the dispersion of the peak distribution over them allows us to derive error bars that takes into account sample variance. These error bars are in very good agreement with those of the covariance matrix, as can be seen from Fig.~\ref{fig:distripeak} where bootstrap errors are represented in magenta and those from the simulations in green, highlighting that the simulations are a good representation of the data.

Because the peak distribution is dominated by noise, we need to run several realisations of the observed shape noise so that the simulations are not biased to one particular realisation of shape noise. For every simulation we run 5 random noise realisations. We also build 5 random noise-only peak distributions from the observations by computing the aperture mass map with all galaxies being randomly rotated. Each of these 5 realisations is computed with a different random seed but the seed is the same for all different cosmologies and for the noise-only realisation, limiting the impact of random shot noise. The simulations at the fiducial cosmology all have different random seeds because they are used to estimate the covariance matrix. This allows us to measure differential peak counts, i.e. the peak distribution in the aperture mass map from which we subtract the distribution of noise peaks. We verified that increasing the number of realisations to 20 does not affect the cosmological constraints (less than 0.3$\sigma$ change).

Figure~\ref{fig:distripeak} shows the main results concerning the peak distribution. It displays the peak distribution, the noise distribution, and the differential distribution for the observation and for all simulations. Simulated peak distributions are the mean over the different noise realisations. We see in particular that the peak distribution is dominated by shape noise, but that we can control it by having the same noise in the data and the simulations. Looking at the differential peak distribution we see a good agreement between the data and the simulations with $S_8$ slightly higher than the fiducial cosmology. We note also that the simulated peak distributions vary smoothly with cosmology with an increasing number of high S/N peaks when $S_8$ is higher. This is expected: an increase in $\Omega_{\rm m}$ increases the mass content of the Universe and an increase in $\sigma_8$ increases the clustering of structures, which both lead to more massive halos and therefore more high-S/N peaks. In the low-S/N regime we note that the differential peak distribution gets negative. This is because the aperture mass distribution is a convolution between signal which presents a high-S/N tail and random Gaussian noise. The noise acts as a Gaussian smoothing and lowers the amplitude in the convex parts of the distribution while it increases it in the concave parts. The peak distribution is thus of higher amplitude than the noise-only distribution at high S/N, of lower amplitude at low S/N, and of equal amplitude when the second derivative of the peak distribution is equal to zero, here around ${\rm S/N}=2$. This argument is rigorously true only for the aperture mass distribution but we extend it to the peaks as they are a subsample of the latter distribution. Finally, we note that the observed differential number of peaks deviates by more than $2\sigma$ from the expectation at S/N $=1.33$ and S/N $=2.33$. Dividing the data in several subareas, we find that only some of the patches are affected by these offsets, but could not find an obvious cause to them. This plot shows that the peak distribution is sensitive to cosmology and that the combination of the KiDS observations and of the simulations we are using does enable us to constrain the cosmological parameter $S_8$. Although it would be tempting to use the differential peak counts for extracting cosmology we prefer to work with the non-subtracted peak distribution to avoid biasing the data which contain only one realisation of the noise.

\subsection{Interpolation}
\label{subsec:interp}

\begin{figure}
\centering
\includegraphics[width=0.7\textwidth,clip,angle=0]{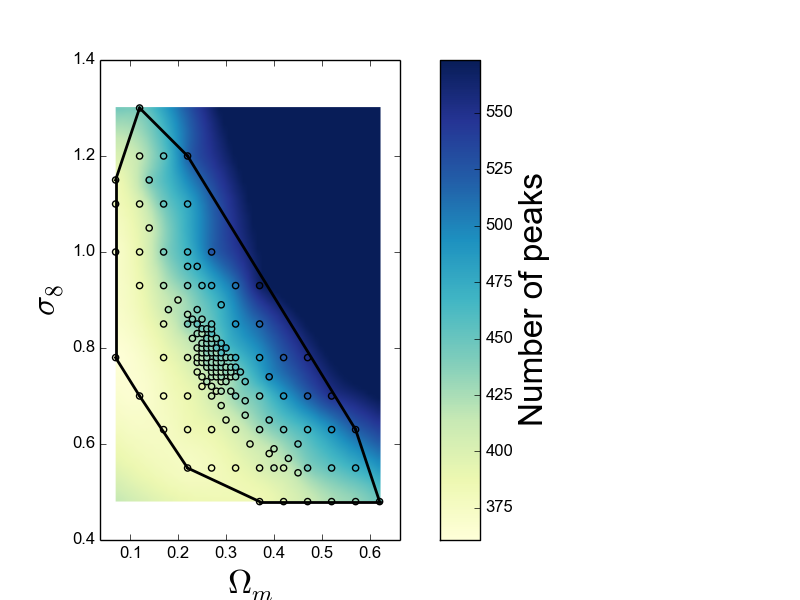}
\caption{Interpolation of the number of peaks in the bin $3.33 \leq {\rm S/N} \leq 3.66$. Dots correspond to measured values and the background area to interpolated ones. The black polygon represents the convex hull within which we trust the interpolation. See text for details.}
\label{fig:peaksint}
\end{figure}

Due to the prohibitive computational cost of simulating a cosmological grid evenly sampling the $(\Omega_{\rm m}, \sigma_8)$ plane, we interpolate the peak distribution at the grid values. We cover a regular grid with step size 0.01 in each direction. Each bin of the data vector is interpolated separately. Peak distributions are averaged over the different ray-tracing and noise realisations before performing the interpolation so that we are not biased by a particular noise realisation. We also recall that shot noise is reduced by applying the same random shape noise to all cosmologies (but the fiducial) for each noise realisation. We note that it is also possible to directly interpolate the likelihood instead of the peak distribution, but the former method is preferable as it interpolates the expected values of the peak distribution while the latter also affects the data vector which enters the likelihood.

We use radial basis functions with a multiquadric model which renders well the evolution of the number of peaks with $\Omega_{\rm m}$ and $\sigma_8$ \citep{Liu+15j}. The computation is performed through the {\it scipy.interpolate.Rbf} Python function (\url{https://docs.scipy.org/doc/scipy/reference/generated/scipy.interpolate.Rbf.html}).

Because the variation of the peak distribution with cosmological parameters is noisy we also add some smoothing when interpolating the peak distribution through the ``smooth'' argument of the {\it scipy.interpolate.Rbf} function  which reduces the number of nodal points in the interpolation process. This improves the rendering of the significance contours in the $(\Omega_{\rm m}, \sigma_8)$ plane and we verify that it does not affect the estimated value of $S_8$. We also check that the error on the interpolated number of peaks is lower than the Poisson error by comparing the results of the interpolation with the measurements for every available simulation (see Appendix~\ref{app:interp} for details). In principle we could avoid the smoothing by running simulations for more points in the $(\Omega_{\rm m}, \sigma_8)$ plane, but this would be computationally demanding, and unnecessary as we found that the constraints on $S_8$ do not change for various values of smoothing. This would however improve the cosmological contours in 2D-space. \citet{Zorrilla+16} also found that using a refined grid of $\sim8000$ points in the $(\Omega_{\rm m}, \sigma_8)$ plane instead of $\sim160$ impacts the tails of the 2D contours but not seriously so in the direction perpendicular to the main degeneracy, so that the undersampling does not significantly affect the estimate of $S_8$.

An example of the interpolated number of peaks in the $(\Omega_{\rm m}, \sigma_8)$ space is given in Fig.~\ref{fig:peaksint}. We see that the interpolation performs reasonably well comparing interpolated values to the nearby measured data points (see Appendix~\ref{app:interp} for the quantitative comparison). However, the extrapolation is very inaccurate. We therefore apply a prior on the likelihood to discard the extrapolation region. This region is defined through a convex hull on the ensemble of points where simulations were run, and is displayed in Fig.~\ref{fig:peaksint}.

\section{COSMOLOGICAL CONSTRAINTS}
\label{sec:cc}

\subsection{Inferring cosmological parameters}
\label{subsec:cosmonoise}

Cosmological parameters are estimated by comparing the observed peak distribution to that of simulations with various cosmologies, in a Bayesian framework.

Our data vector is represented by $\bmath{x}\in\mathbb{N}_+^{N_{\rm b}}$, the number of peaks in each of the $N_{\rm b}$ bins of S/N shown in Fig.~\ref{fig:distripeak}. Similarly we define the peak distribution of a simulation with cosmology $\bmath{\pi}=(\Omega_{\rm m}, \sigma_8)$ as $\bmath{x_{\rm s}}(\bmath{\pi})\in\mathbb{R}_+^{N_{\rm b}}$.

From Bayes theorem we can link the probability $p(\bmath{\pi}|\bmath{x})$ of one cosmological model given the data vector (i.e. what we want to know) to the probability $p(\bmath{x}|\bmath{\pi})$ of the data vector given a cosmology;

\begin{equation}
p(\bmath{\pi}|\bmath{x})=\frac{p(\bmath{x}|\bmath{\pi})p(\bmath{\pi})}{p(\bmath{x})}.
\end{equation}

\noindent The probability of the data $p(\bmath{x})$ is a normalisation constant and $p(\bmath{\pi})$ is a flat prior, with value 1 on the probed range of cosmologies $\bmath{\pi}$ within the convex hull shown in Fig.~\ref{fig:peaksint}, and 0 elsewhere.

The peak distribution is assumed to be a multivariate Gaussian distribution. This approximation is valid provided that we have a sufficient number of peaks in each bin, typically a few tens. The main analysis is done with 12 bins evenly spaced between S/N of 0 and 4 but we also check the robustness of our results over two alternative bin widths (0.25 and 0.5). We do not use a Gaussian likelihood but the adapted version of a multivariate $t$-distribution presented in \citet{Sellentin+16}, which still assumes Gaussian distributed data. This likelihood, derived from marginalising over the true covariance matrix, provides better inference than the traditional Gaussian likelihood with the \citet{Hartlap+07} correction, which only gives an unbiased estimate for the inverse covariance matrix. The likelihood can be written as

\begin{equation}
\label{eq:bay}
p(\bmath{x}|\bmath{\pi}) = c(N_{\rm s},N_{\rm b})\sqrt{\det\Sigma(\bmath{\pi})} \left[ 1+\frac{\chi^2 \left(\bmath{x},\bmath{\pi}\right)}{N_{\rm s}-1}\right]^{-N_{\rm s}/2},
\end{equation}

\noindent where $N_s$ is the number of simulations used to estimate the covariance matrix $\Sigma$, $c(N_{\rm s},N_{\rm b})$ is a constant which depends on the number of simulations and the size of the data vector, and $\chi^2 \left(\bmath{x},\bmath{\pi}\right)$ is the $\chi^2$ function defined in Eq.~(\ref{eq:chi2}). We note that this likelihood approaches a Gaussian likelihood when the number of simulations $N_{\rm s}$ is large.

With the assumption that the covariance matrix does not depend on $\bmath{\pi}$, the numerator of Eq.~(\ref{eq:bay}) is constant and we can write

\begin{equation}
p(\bmath{x}|\bmath{\pi}) \propto \left[ 1+\frac{\chi^2}{N_{\rm s}-1}\right]^{-N_{\rm s}/2},
\end{equation}

\noindent where the covariance matrix is computed from the $N_{\rm s}=175$ simulations of the fiducial cosmology $\bmath{\pi_0}$,

\begin{equation}
\Sigma(\bmath{\pi_0}) = \frac{1}{N_{\rm s}-1}\sum_{i}^{N_{\rm s}}{(\bmath{x_{{\rm s},i}}(\bmath{\pi_0})-\bar{\bmath{x_{\rm s}}}(\bmath{\pi_0})) \; (\bmath{x_{{\rm s},i}}(\bmath{\pi_0})-\bar{\bmath{x_{\rm s}}}(\bmath{\pi_0}))^{\rm T}}.
\end{equation}

\noindent The vector $\bmath{x_{{\rm s},i}}(\bmath{\pi_0})$ represents the peak distribution of the $i$-th fiducial simulation and $\bar{\bmath{x_{\rm s}}}(\bmath{\pi_0})$ is the mean peak distribution over all fiducial simulations.

Because we do not have the computational resources to compute the covariance matrix at each cosmology, we make the assumption that it does not depend on cosmology. Although this approximation does not hold for large variations in the cosmological parameters, \citet{Eifler+09} showed that it overestimates the errors on cosmological parameters in the case of 2nd-order cosmic shear (which contains overlapping information with peaks) such that our constraints are conservative. We also note that \citet{Zorrilla+16} found a $\sim 15\%$ improvement in the cosmological parameter forecasts from peaks in simulations when taking into account the cosmological dependence of the covariance matrix, such that the constraints presented in this paper could be further improved with extra simulations at the non-fiducial cosmologies.

The covariance matrix estimates the error correlations in the data. The main sources of errors are galaxy shape noise and sample variance. The first one is probed by applying different random orientations to the intrinsic ellipticities of galaxies, and the second one by using several simulations and several ray-tracings through the simulations. \citet{Kacprzak+16} focus on the shape noise contribution by applying many realisations of shape noise to the same simulations. This approach allows them to have a higher number of data vectors in the covariance matrix computation but neglects the contribution from sample variance over that of shape noise. In contrast, we estimate our covariance matrix with $N_{\rm s}=175$ independent data vectors from 35 different simulations with 5 different ray-tracing each, and different shape noise realisations. We compute 5 covariance matrices with different seeds for shape noise and average the covariances. We use this approach because we find that shape noise and cosmic variance affect the peak distribution at the same level. The peak distribution of 10 fiducial different simulations with the same shape noise and that of one fiducial simulation with 10 different realisations of the shape noise represents a dispersion of the same order, typically a few to ten percent of the mean value. With this strategy we estimate an accurate covariance matrix without biasing with non-independent data vectors. The cosmological constraints are almost identical for any individual matrix, but using the average covariance avoids choosing one set of noise realisations over another.

The $\chi^2$ is defined in Eq.~(\ref{eq:chi2}) from comparing the observed data vector $\bmath{x}$ to the model $\bmath{x_{\rm s}}(\bmath{\pi})$ estimated from a simulation with cosmological parameters $\bmath{\pi}$, using the covariance matrix evaluated at the fiducial cosmology $\bmath{\pi_0}$:

\begin{equation}
\label{eq:chi2}
\chi^2(\bmath{x},\bmath{\pi}) = (\bmath{x}-\bmath{x_{\rm s}}(\bmath{\pi}))^{\rm T} \; \Sigma^{-1}(\bmath{\pi_0})  \; (\bmath{x}-\bmath{x_{\rm s}}(\bmath{\pi}))
\end{equation}

In contrast to the case of 2PCFs, there is no simple analytical prescription for the variation of the peak distribution with cosmology $\bmath{x_{\rm s}}(\bmath{\pi})$. In fact analytical models exist for the high-S/N peaks, as used in \citetalias{Shan+17}, but cannot be applied to lower-S/N peaks. For each cosmology we therefore average the peak distribution over the different realisations of cosmic variance and shape noise, before using them in the $\chi^2$ computation. We note that the goal here is to have the best knowledge of the expectation value which is different than in the covariance matrix where we want to estimate the variation of the peak distribution with noise. This is also the reason why for the different cosmologies we use the same noise seeds but not for the fiducial ones. Using different seeds for shape noise would increase shot noise between the different cosmologies, requiring to average over a larger number of realisations to extract the cosmological dependence of the peak distribution.

The likelihood $p(\bmath{x}|\bmath{\pi})$ is computed at each point of the interpolated grid of parameters, and normalised by the integrated likelihood over the prior support. We then determine the 1$\sigma$ (resp. 2$\sigma$) iso-likelihood contours as the contours enclosing 68\% (resp. 95\%) of the total integrated likelihood. For each parameter we also estimate the most favoured value as the maximum of the likelihood marginalised over the other parameter, and the 1$\sigma$ uncertainty such that it encloses 68\% of the marginalised likelihood integrated over the probed parameter range. As the likelihood is computed in the $(\Omega_{\rm m}, \sigma_8)$ plane we apply a change of variables to measure constraints on $S_8=\sigma_8 \sqrt{\Omega_{\rm m}/0.3}$:

\begin{equation}
p(\Omega_{\rm m},S_8|\bmath{x}) \; {\rm d}\Omega_{\rm m} \; {\rm d}S_8 = p(\Omega_{\rm m},\sigma_8|\bmath{x}) \; {\rm d}\Omega_{\rm m} \; \frac{\partial S_8}{\partial \sigma_8}\bigg\rvert_{\Omega_{\rm m}} \; {\rm d}\sigma_8.
\end{equation}

\subsection{Systematics}
\label{subsec:sys}

Cosmological constraints from shear peak statistics are affected with several systematics, namely: multiplicative shear bias, mean redshift bias, baryon feedback, intrinsic alignment, and boost factor. Although the impact of these biases on convergence peaks has been discussed in detail in \citetalias{Shan+17}, they might affect the present analysis differently due to the different methodology and using low-S/N peaks.

\citet{Hildebrandt+17} found that the multiplicative shear bias and mean redshift bias only have a small impact on $S_8$, in the case of 2PCFs applied to KiDS-450. In addition, \citet{Kacprzak+16} also found almost no impact on $S_8$ central values in the case of peak statistics in the DES-SV data, and that neglecting these biases leads to 15\% tighter constraints with the same definition of $S_8$ as in our paper. However, in KiDS-450 we have redshift bias better than $\Delta z \sim 0.02$ \citep{Kuijken+15} against $\sim 0.05$ in DES-SV, and a shear multiplicative bias $m \lesssim 0.01$ \citep{FenechConti+17} against $\sim 0.05$ in DES-SV, such that these two biases should be smaller in the present study than in \citet{Kacprzak+16}. In the case of KiDS-450 2PCFs the multiplicative shear and photometric redshift biases have negligible effects on the $S_8$ value and present uncertainties of about 1.7\% and 0.8\%, respectively. Assuming that these biases impact peak statistics at the same level as they impact the 2PCFs, we can derive conservative constraints by adding a null bias to our $S_8$ estimate and adding the uncertainty on these biases in quadrature to the statistical error. In principle it is possible to account for $m$/$\Delta z$ biases by modifying their values in the simulations, computing the dependency of the peak distribution on these biases, and then marginalize over it. However this would require lots of computational time for such a small bias as noted in \citet{Hildebrandt+17}.

The \citet{DH10} simulations are DM only, such that they neglect the impact of baryons, which can modify how LSS evolves. Using a set of hydrodynamical simulations, \citet{Osato+15} measured the impact of baryons on both power spectrum and peak statistics. Their simulations also account for feedback from supernovae and active galactic nuclei. They find a similar bias due to baryonic effects for different ranges of peak amplitudes between $1\leq {\rm S/N} \leq5$, and estimate an $\sim1.5$\% effect on both $\Omega_{\rm m}$ and $\sigma_8$. This propagates to a $-2.3\%$ effect on $S_8=\sigma_8 \sqrt{\Omega_{\rm m}/0.3}$. We note that we also include peaks in the range $0\leq {\rm S/N} \leq1$ in our analysis, but \citet{Yang+11} found that for low peaks, the impact of baryons is mitigated, because light-rays towards the peaks do not pass near the cores of the halos along the line of sight, such that our estimate of the baryon bias should be conservative. Our filter function also down-weights the central part, i.e., the peak height is not primarily determined by the central portion of the halo if the matched filter and the halo are aligned. This further decreases the impact of baryonic effects. We apply a $-2.3\%$ bias to our $S_8$ estimate in order to correct for baryons. We also add this value in quadrature to the error budget, therefore assuming an uncertainty on the bias as large as the bias itself. This is a conservative approach to account for the fact that we do not accurately know the uncertainty on this bias.

On small scales IA refers to the radial alignment of satellite galaxies within DM halos which breaks the fundamental assumption of WL that galaxies are randomly oriented. These alignments are generated by the gravitational potential of high-mass halos on neighbouring galaxies. This effect is divided into two components: the intrinsic-intrinsic correlations (II), i.e. the alignment of galaxies physically linked together, and gravitational-intrinsic correlations (GI), i.e. the alignment of halo galaxies with the induced shear on background galaxies \citep{Hirata+04}. In the case of peaks, the effects of IA can be captured by modelling the alignment of satellite galaxies towards DM halo centres, e.g., the \citet{Schneider+10} model. Using this model with the fiducial value for the alignment strength prescribed in \citet{Schneider+10}, \citet{Kacprzak+16} found a change in the amplitude of shear peaks lower than 5\%, applying the same methodology as ours to the DES-SV data. We also note that \citet{Sifon+15} measured the radial alignment of satellite galaxies in a sample of 90 galaxy clusters, securing cluster membership through spectroscopic redshifts, and found negligible alignments. Based on their measurement they show that the \citet{Schneider+10} recommended alignment strength overestimates the IA at small scales (see their Fig. 13), such that the effect of IA on the peak distribution is probably much lower than what \citet{Kacprzak+16} found.

The dilution of the background shear signal due to the inclusion of cluster galaxies is generally compensated for by a radially-dependent boost factor to the shear in cluster lensing studies \citep[e.g., ][]{Applegate+14,Hoekstra+15,Martinet+16}. In the case of peak statistics the contamination from cluster galaxies leads to higher peaks in the simulations than in the observations. Around an observed galaxy cluster, the background shear signal is diluted. But in the simulated mocks where galaxies have the same positions as in the data, at a DM halo position there is no dilution of the shear signal because the distribution of galaxies is imposed by the data. Comparing the radial profile of galaxy density at peak locations in the observation with that of simulations allows one to compute the boost factor in bins of peak S/N. This also accounts for the loss of correlation between the lensing mass and source galaxy distributions when adapting the simulated mocks to the observation (see Sect.~\ref{subsec:sims:adapt}). With the same peak calculation and simulations as ours, \citet{Kacprzak+16} estimated the variation of the number of peaks per S/N bin due to the boost factor in the DES-SV. They found a variation which is proportional to the S/N of peaks and lower than about 5\% for S/N lower than 4, and therefore recommend using bins with S/N lower than this value to avoid large shear dilution effects. In a similar approach but on convergence peaks in KiDS-450, \citetalias{Shan+17} found a change of about 6\% and 10\% in the number of peaks in the bin with $3<{\rm S/N}<3.5$ and $3.5<{\rm S/N}<4$ respectively which correspond to the highest S/N used in this study, and is comparable with the results from DES-SV although the redshift distributions of both surveys are different.

Applying both IA and boost factor corrections, \citet{Kacprzak+16} found a variation of $S_8$ of 0.01 using shear peaks defined with the same filter as ours, corresponding to a systematics relative bias of $\sim 1.3\% $. We note that IA and the boost factor tend to increase $S_8$ together. Based on the discussion of the two last paragraphs we can assume this value to be an upper limit for this systematic bias in the case of KiDS-450. We add the above estimate to our $S_8$ value and add it in quadrature to the error budget. As noted in \citet{Kacprzak+16}, current models correcting for IA and boost factors have a high uncertainty in the case of peak statistics. This highlights a lack of extensive study on the impact of these systematics on peak statistics, and dedicated studies are required to improve these models, which is beyond the scope of this paper.

The biases estimated above are linearly added to our $S_8$ best estimate and the uncertainties on these biases are added in quadrature to the statistical 68\% errors on $S_8$. We note that except for the multiplicative shear and mean redshift biases for which we have estimates of the uncertainties, we assumed that the uncertainty on each bias is as large as the bias itself. This allows us to correct for biases in a conservative maner although we lack precise information on the bias uncertainties in the case of baryons, IA, and boost factor. In doing so we also neglect any correlation between the different systematics, except that between the boost factor and IA which are treated together. The joint contribution of every bias leads to a shift of the $S_8$ value of -0.95\%, which is lower than the percent because some biases compensate each other. The total systematic uncertainty is $\sim3.2\%$ and is dominated by baryon feedback. It is added in quadrature to the statistical precision. This value is also similar to the $\sim 3.6\%$ systematic uncertainty that was assigned to the 2PCFs analysis \citep{Hildebrandt+17}.

\subsection{Results}
\label{subsec:res}

\begin{figure}
\centering
\includegraphics[width=0.5\textwidth,clip,angle=0]{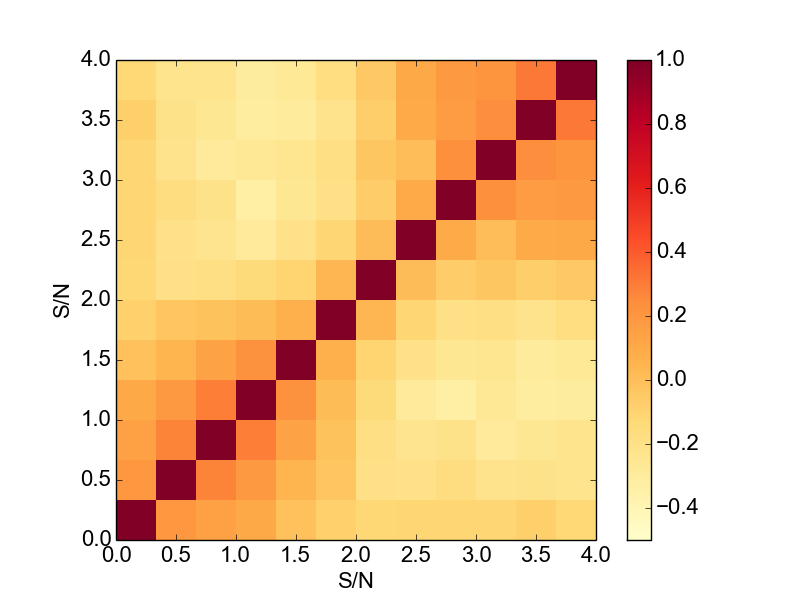}
\caption{Correlation of the covariance matrix for peaks in the range $0\leq {\rm S/N} \leq 4$.}
\label{fig:corcov}
\end{figure}

We first show the correlation matrix in Fig.~\ref{fig:corcov}. As mentioned earlier, we work with the mean covariance matrix over 5 realisations of shape noise, decreasing shot noise in the covariance estimate, although it is still representative of the noise in the data. We note that low-S/N peaks are slightly correlated with one another (${\rm S/N} < 1.66$), and high-S/N peaks (${\rm S/N} > 2.66$) show even stronger correlations. This is expected as a massive halo tends to correspond to several peaks both due to its large size and its large amount of substructures. However, we see only small correlations between the two regimes of peaks, with close to zero negative off-diagonal terms. This means that the low- and high-S/N peaks probe different information, projections of small structures and high-mass halos, respectively. The slight anti-correlation in between the two regimes is also present in the covariance matrix of \citet{Zorrilla+16} and is due to the fact that when a large halo is detected, projection effects around this halo fade. This is also seen in the peak distribution (Fig.~\ref{fig:distripeak}) which shows negative differential peak counts in the low-S/N regime and positive ones in the high regime. These ranges of S/N also roughly correspond to the S/N where the peak distribution is the most sensitive to cosmology, as seen in Fig.~\ref{fig:distripeak}.

\begin{figure}
\centering
\includegraphics[width=0.5\textwidth,clip,angle=0]{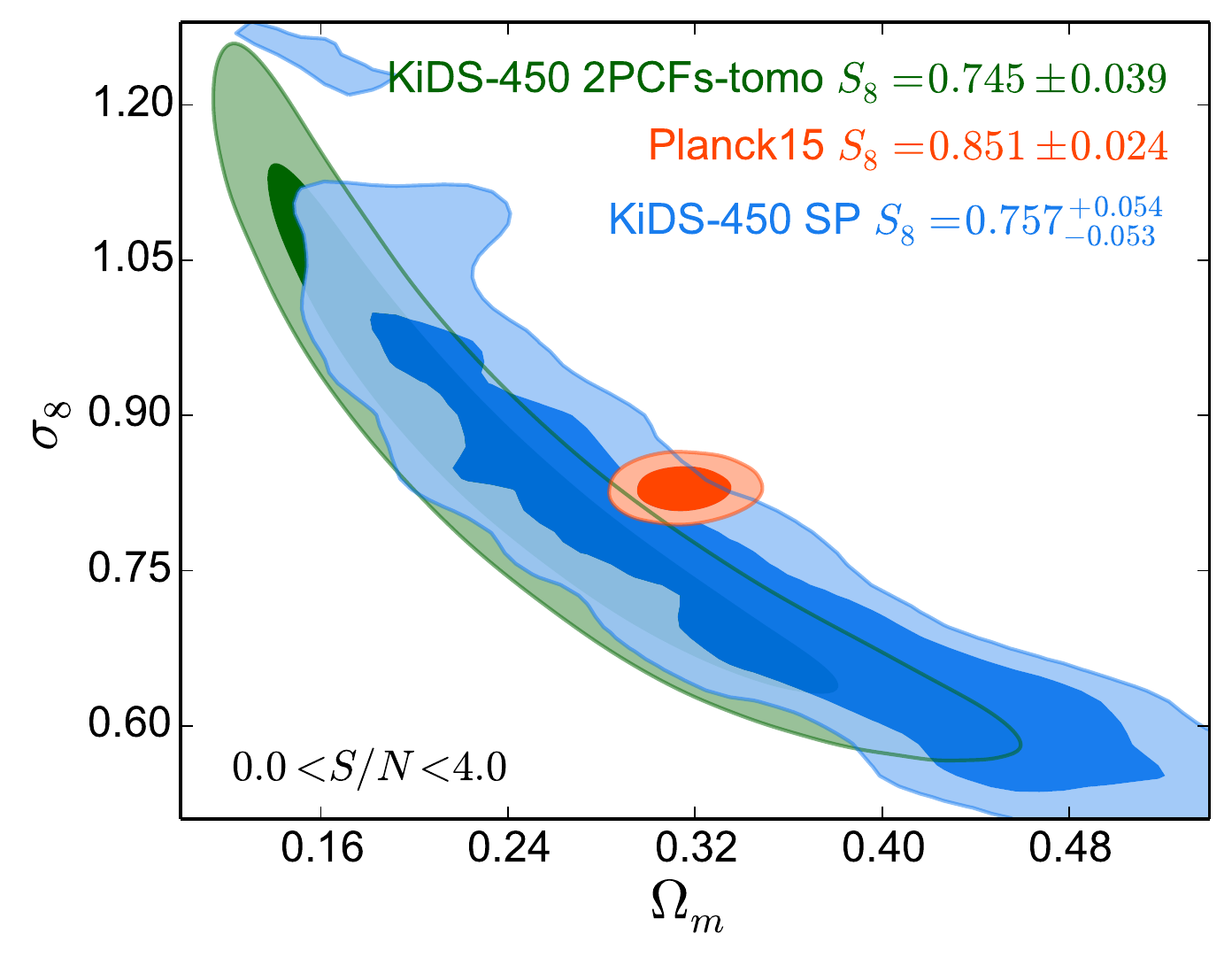}
\includegraphics[width=0.5\textwidth,clip,angle=0]{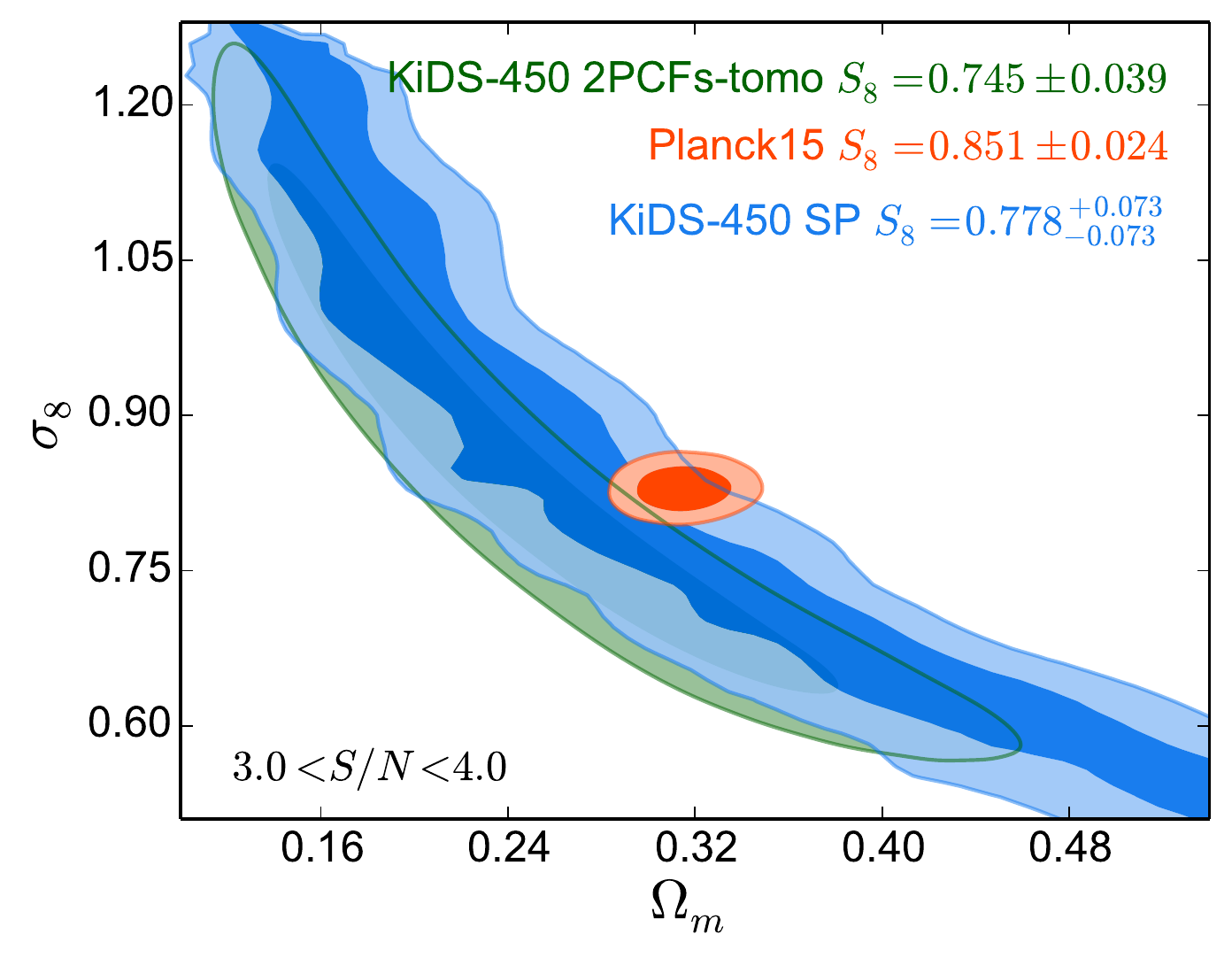}
\caption{Cosmological constraints on ($\Omega_{\rm m},\sigma_8$). 1$\sigma$ and 2$\sigma$ confidence regions are displayed in dark and light areas, respectively. Green contours correspond to KiDS-450 tomographic 2PCFs constraints marginalised over systematics \citep{Hildebrandt+17}, blue to the KiDS-450 shear peak constraints using bins in ranges $0 \leq {\rm S/N} \leq 4$ ({\it top}), and $3 \leq {\rm S/N} \leq 4$ ({\it bottom}), and red to Planck ``TT+lowP'' \citep{PlanckXIII}. The top right legend shows the $S_8$ best value and 68\% errors for each study.}
\label{fig:cosmores}
\end{figure}

Cosmological constraints from shear peak statistics are displayed in Fig.~\ref{fig:cosmores}, where we show the 1 and 2$\sigma$ contours for the 2D likelihood, and the $S_8$ best estimate from the marginalised 1D likelihood. We do not present any estimate of $\Omega_{\rm m}$ or $\sigma_8$ because they are highly correlated as shown by the large degeneracy in Fig.~\ref{fig:cosmores}. We present constraints using the full range of available peak S/N ($0\leq {\rm S/N} \leq 4$), and also using only the high S/N peaks ($3\leq {\rm S/N} \leq 4$). This second plot serves to assess the gain of information from the low-S/N peaks, and also to allow a comparison with peak constraints from analytical predictions as in \citetalias{Shan+17}. We also note the presence of wiggles in the contours, which are an artefact of the interpolation of the peak distribution with large separation between points in the $(\Omega_{\rm m}, \sigma_8)$ plane. These wiggles would disappear if we could use simulations paving more points in the cosmological parameter space. Our best estimates are $S_8=0.757^{+0.054}_{-0.053}$ (68\% errors) for the full range of S/N, and $S_8=0.778^{+0.073}_{-0.073}$ when focusing on high S/N only. Including the systematics estimated in Sect.~\ref{subsec:sys} yields $S_8=0.750^{+0.059}_{-0.058}$ and $S_8=0.771^{+0.077}_{-0.077}$ for all and high-only S/N, respectively. The statistical error on $S_8$ is $\sim7.1\%$ (resp. 9.4\% for high-only peaks) and the systematic uncertainty is $\sim3.2\%$. Statistical errors therefore dominate systematic ones in the case of KiDS-450. This will no longer be the case for larger surveys and detailed studies are required to better understand, and correct for the systematics affecting shear-peak statistics.

In Appendix~\ref{sec:discussion:cov}, we make use of the refined SLICS simulations to verify that the assumptions made in the case of the \citet{DH10} simulations do not significantly affect the main results of the paper. We find that the refined covariance matrix computed from the SLICS simulations present similar correlations as that of the fiducial mocks, but a higher scatter due to a better inclusion of sample variance. With the refined covariance matrix we find $S_8=0.760^{+0.061}_{-0.058}$ and $S_8=0.771^{+0.074}_{-0.075}$ respectively for $0 \leq {\rm S/N} \leq 4$ and $ 3 \leq {\rm S/N} \leq 4$, and with accounting for systematics. The constraints on $S_8$ are left almost unchanged by switching between the original and the refined covariance matrix, validating the various approximations made in the \citet{DH10} mocks (e.g., interpolation, redshift range). We also note that the degeneracy in the ($\Omega_{\rm m}$, $\sigma_8$) plane does not change. Although the sample variance bias of our simulations has negligible effect on the present study, it will become more important for larger area surveys and it might become necessary to use simulations which cover an area which is close to that of the data to account for sample variance.

\section{DISCUSSION}
\label{sec:discussion}

\begin{figure*}
\centering
\includegraphics[width=1.0\textwidth,clip,angle=0]{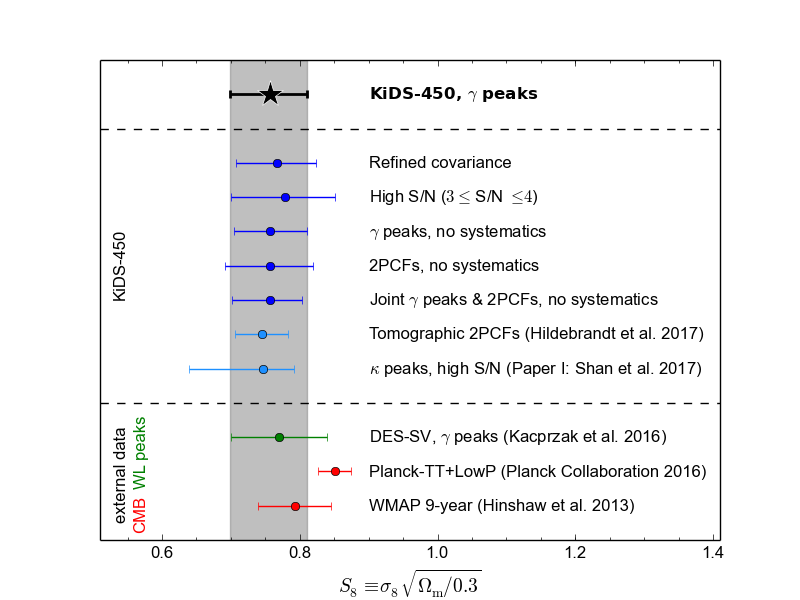}
\caption{Comparison between various constraints on $S_8$. The black star represents our best estimate, and the shaded area its error bars. Dark and light blue points correspond to different constraints with peak statistics and 2PCFs, from this paper and from other KiDS-450 papers respectively. The green point corresponds to a shear peak study for another dataset, and red points show constraints from CMB measurements.}
\label{fig:comp}
\end{figure*}

Figure~\ref{fig:comp} summarises $S_8$ constraints from this survey and compares them with various other studies. We calculated $p$-values as an estimate for the goodness-of-fit for all the cases considered. The $p$-values are calculated for the minimum $\chi^2$ taking into account the degrees-of-freedom given by the number of data points minus two free parameters ($\Omega_{\rm m}$ and $\sigma_8$). All the values are larger than 0.2, indicating that the models fit the data well.

\subsection{Information from low- and high-S/N peaks}

We first focus on the gain of information from adding the low-S/N peaks. We recall that the large-S/N peaks correspond to single massive halos while the low-S/N correspond to alignment of smaller halos along the line-of-sight. We find very good agreement between the two regimes, showing that chance alignments and larger halos are both good tracers of LSS. The constraints shrink by 24\% when adding the low-S/N peaks, representing a large gain of information. This highlights the great interest of studying the low-amplitude peaks, which efficiently probe the cosmological information contained in the chance alignments of LSS. This result is also supported by the study of \citet{Shirasaki17} who showed that low-amplitude peaks contain significant non-Gaussian information.

In Appendix~\ref{app:negpeaks}, we additionally probe the potential gain from including negative S/N peaks. However, we found that those do not further increase the constraints on $S_8$ compared to the positive peaks only. This can be explained by the high correlation that we find between the negative and positive peaks.

\subsection{Comparison with KiDS 2PCFs and Planck}

One of the goals of this study is to check whether peak statistics agree with KiDS 2PCFs, in light of the reported mild tension between the latter and Planck results.

Peak statistics yield similar constraints on $S_8$ as 2PCFs. In particular the degeneracy in the ($\Omega_{\rm m}, \sigma_8$) plane is parallel to that of 2PCFs (Fig.~\ref{fig:cosmores}), highlighting the strong correlation between the two probes. We note that our estimate of $S_8$ is in good agreement with the tomographic 2PCFs value reported in \citet{Hildebrandt+17}: $S_8=0.745\pm0.039$. We stress, however, that in the case of peaks we included the different systematics (multiplicative shear bias, mean redshift bias, baryon feedback, intrinsic alignment, boost factor) as a correction to the best estimate and uncertainties of $S_8$ while \citet{Hildebrandt+17} marginalised over the relevant systematics. \citet{Hildebrandt+17} also varied all cosmological parameters while we can only vary $\Omega_{\rm m}$ and $\sigma_8$ in the case of peaks with the given set of simulations, and therefore underestimate the confidence regions. Finally, we note that we did not carry out a tomographic analysis of the peak statistics, while the 2PCFs study captures the information from four different source redshift bins. This choice is due to limitations in the available simulation mocks, and explains why constraints are tighter in the case of 2PCFs. \citet{Martinet+15b} showed that a tomographic approach can improve constraints from peak statistics by almost a factor of two in the case of {\it Euclid}-like simulations. The improvement from a tomographic peak analysis has also been noted by \citet{DH10} using CFHT-like simulations, and by \citet{Petri+16} with LSST-like simulations.

We find only a slight difference when comparing peak statistics with Planck CMB. Our constraints on $S_8$ present a 1.6$\sigma$ difference with that of Planck \citep[$S_8=0.851\pm0.024$, ][]{PlanckXIII}, when including systematics. \citet{Hildebrandt+17} reported a tension of 2.3$\sigma$, however our constraints are $33\%$ weaker in comparison, mainly due to the fact we are not using tomography: we show in Sect.~\ref{sec:discussion:joint} that peak statistics and 2PCFs achieve similar constraints when tomography is not used in both cases. Our best estimate of $S_8$ is however closer to that of \citet{Hildebrandt+17}. Finally, we note that our simulations are run with a Hubble parameter $H_0=70$~km~s$^{-1}$~Mpc$^{-1}$ which is different from the Planck estimated value: $H_0=67.3$~km~s$^{-1}$~Mpc$^{-1}$. While this could partially explain the difference between the two probes, it is difficult to assess in this paper, as it would require to run extra simulations fixing $H_0$ to the Planck value. We note that our constraints do not change significantly when using the refined covariance matrix (see Sect.~\ref{sec:discussion:cov}) which is computed with a slightly different Hubble parameter: $H_0=69.0$~km~s$^{-1}$~Mpc$^{-1}$. In addition, we found in \citetalias{Shan+17} that the constraints on $S_8$ are stable when varying $H_0$ between 68~km~s$^{-1}$~Mpc$^{-1}$ and 72~km~s$^{-1}$~Mpc$^{-1}$ with the modeled peak function, showing that the observed difference to Planck is probably not due to different values of the Hubble constant.

\begin{figure*}
\centering
\includegraphics[width=1.0\textwidth,clip,angle=0]{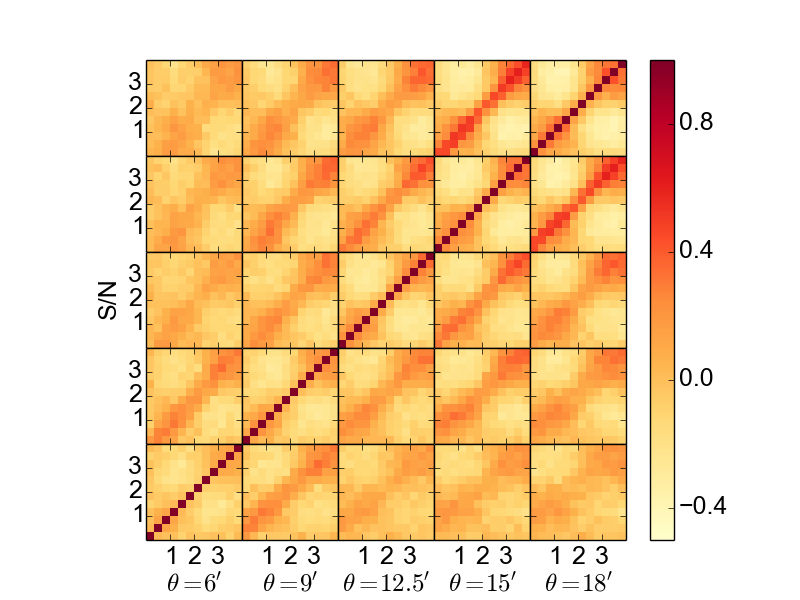}
\caption{Correlation of the multi-scale covariance matrix for peaks in the range $0\leq {\rm S/N} \leq 4$. Each square represents the correlation of the peak S/N distribution between two filter scales $\theta$ in $(6\arcmin, 9\arcmin, 12.5\arcmin, 15\arcmin, 18\arcmin)$.}
\label{fig:corcov5}
\end{figure*}

\subsection{Comparison with other WL peak analyses}

We first compare our constraints to that of \citetalias{Shan+17}, which uses the same KiDS-450 dataset but apply a different approach to peak statistics. \citetalias{Shan+17} uses convergence peaks instead of shears, and also compares their peak distribution to that from analytical predictions, although calibrating those on the same simulations as the ones we use in the present study. \citetalias{Shan+17} uses a Gaussian filter with a scale of 2 arcmin on the convergence field, which is close to the effective radius of our own filter ($\theta=1.875\arcmin$). As the analytical predictions are accurate only in the high-S/N regime where one can assume that peaks are associated with high-mass halos, \citetalias{Shan+17} makes use of the high-S/N peaks only ($3 \leq {\rm S/N} \leq 5$). We therefore only compare our high-S/N ($3 \leq {\rm S/N} \leq 4$) constraints with the results from \citetalias{Shan+17}. In this range of S/N \citetalias{Shan+17} finds $S_8=0.746^{+0.046}_{-0.107}$. We note that these constraints are in good agreement with ours, with the same constraining power as our high-S/N regime (see Fig.~\ref{fig:comp}). The fact that both studies, while based on completely different approaches, give consistent $S_8$ estimates is a good assessment of their robustness.

There is, however, a major difference in the orientation of the ($\Omega_{\rm m}, \sigma_8$) degeneracy. In the present study we find a degeneracy similar to that of the 2PCFs even for the high peaks, while in \citetalias{Shan+17} a flatter degeneracy is seen, closer to that of some cluster studies. The only way for us to reproduce this degeneracy is to use only the highest S/N peaks: $4\leq {\rm S/N}\leq 5$. This would mean that even in the $3\leq {\rm S/N} \leq 4$ range there is a large contribution from peaks corresponding to projection effects of LSS and to noise while peaks at ${\rm S/N} \geq 4$ mainly correspond to high-mass halos. The fact that \citetalias{Shan+17} find a flatter degeneracy even when including peaks at $3 \leq {\rm S/N} \leq 4$ could highlight a difference in the definition of S/N which is global in \citetalias{Shan+17} and local here (the noise is different for each aperture), or show some limits of the model used in \citetalias{Shan+17} which only accounts for high-mass halos and not for projection effects. This reasoning holds only under the assumption that the degeneracy is not dominated by other systematic effects. In the literature, we note that different peak studies find different degeneracy orientations. \citet{Liu+15x} find a degeneracy very close to that of \citetalias{Shan+17}, applying the same predictive model to the Stripe82 data acquired with the CFHT. Using simulations instead of analytical prediction, \citet{Liu+15j} find a degeneracy similar to the convergence power spectrum degeneracy, but also using the full range of S/N convergence peaks. \citet{Kacprzak+16} also find a degeneracy in agreement with that of 2PCFs in DES-SV, using simulations and shear peaks in the full range $0\leq {\rm S/N}\leq 4$. Quantitatively, when we let the $\alpha$ parameter free to vary in $\Sigma_8=\sigma_8(\Omega_{\rm m}/0.3)^\alpha$, we find that the best $\alpha$ value is between 0.54 and 0.61, and that our constraints on $\Sigma_8$ are $\sim 5\%$ tighter than those with $\alpha$ fixed to 0.5. This is in good agreement with the best $\alpha$ values of 0.60 and 0.58 found in \citet{Liu+15j} and \citet{Kacprzak+16} but different from that of \citet{Liu+15x} and \citetalias{Shan+17}: 0.43 and 0.38 respectively. Simulation-based peak analyses therefore tend to show degeneracy closer to 2nd-order cosmic shear than to clusters but also often make use of a larger range of S/N compared to model-based analyses which are in better agreement with the degeneracy from cluster analyses.

We can also compare our results with those of the DES-SV analysis \citep{Kacprzak+16}. This study is very similar to ours in its methodology. In particular they use aperture masses with the same filter function in shear space. They also use the same \citet{DH10} simulations to compute their cosmological constraints. Although they define their fiducial results based on a 20 arcmin filter scale, they also provide $S_8$ estimates for a 12 arcmin scale, very close to the 12.5 arcmin used in the present study. We display their value for $\theta=20\arcmin$ in Fig.~\ref{fig:comp} as it corresponds to their fiducial estimate. They find $S_8=0.76\pm0.074$ and $S_8=0.73\pm0.081$ with a $20\arcmin$ and $12\arcmin$ filter scale respectively. These two results are in very good agreement with ours. The error bars are $21\%$ narrower (resp. $28\%$ for $\theta=12\arcmin$) in the case of KiDS due to its larger area (450~deg$^2$ against 139~deg$^2$) and higher galaxy density (8.53 galaxies per square arcmin against 5.7). We also note that \citet{Kacprzak+16} marginalised over the estimated mean redshift and multiplicative biases. While we do not marginalise over systematics, we include them as an a posteriori correction to our best estimate and uncertainties, also accounting for baryon feedback, boost factor and IA. Finally, we note that they found similar $S_8$ when using different aperture sizes. In Sect.~\ref{sec:discussion:multi} we measure the peak statistics for different filter scales and derive constraints from a multi-scale analysis.

The cosmological constraints from \citet{Liu+15x} and \citet{Liu+15j} are not presented in Fig.~\ref{fig:comp} because they use a different definition of $\Sigma_8$ than ours. While we use $S_8=\sigma_8(\Omega_{\rm m}/0.3)^{0.5}$, these two other studies respectively use $\Sigma_8=\sigma_8(\Omega_{\rm m}/0.27)^{0.43}$ and $\Sigma_8=\sigma_8(\Omega_{\rm m}/0.27)^{0.60}$. When recomputing our constraints with their definition we find good agreement with both studies. \citet{Liu+15x} find $\Sigma_8=0.82\pm0.03$ from the $\sim130$ deg$^2$ of the CFHT Stripe 82 Survey with a model-based parameter inference while we find $\Sigma_8=0.768^{+0.063}_{-0.062}$ and $\Sigma_8=0.775^{+0.058}_{-0.057}$ with and without taking systematics into account. \citet{Liu+15j} find $\Sigma_8=0.84^{+0.03}_{-0.04}$ from the $154$ deg$^2$ of the CFHTLenS in the redshift range $0.2<z<1.3$ with a simulation-based parameter inference while we find $\Sigma_8=0.821^{+0.056}_{-0.066}$ and $\Sigma_8=0.829^{+0.049}_{-0.060}$ with and without taking systematics into account. We note that their constraints are tighter than ours while we probe three times as many galaxies as them. \nic{It is hard to assess the origin of this difference, and in particular whether it is due to the different surveys or to the different methods. On the first point we note that in the CFHTLenS they use higher-redshift source galaxies that have a stronger lensing signal compared to lower-redshift galaxies in KiDS. We are currently at an early stage of peak statistics and it would be valuable to apply different techniques to the same datasets to do robust comparisons between methods. We also note that we understand better the constraints when applying the same method to two different surveys as in the case of the comparison between the present analysis and the DES-SV peak study of \citet{Kacprzak+16}.}

\subsection{Multi-scale analysis}
\label{sec:discussion:multi}

We investigate the gain of information from combining the peak statistics of different filter scales. As different filter scales probe different structures, i.e. different halo sizes, combining several scales should yield more precise constraints. However, the information from different scales is correlated: for example a galaxy cluster detected at a smaller scale will be detected at larger scale providing that the scale is not so large that the cluster signal is buried in the noise. See also \citet{Marian+12} for an approach with a single scale of adaptive size. In addition to our fiducial scale of $\theta=12.5\arcmin$, we measure the peak distribution for four extra scales leading to the following ensemble of probed scales: $\theta=(6\arcmin, 9\arcmin, 12.5\arcmin, 15\arcmin, 18\arcmin)$, which respectively correspond to the effective scales $\theta_{\rm eff}=(0.9\arcmin, 1.35\arcmin, 1.875\arcmin, 2.25\arcmin, 2.7\arcmin)$, for the filter parameter $x_{\rm c}=0.15$ (see Eq.~(\ref{eq:Q})). We recall that the fiducial scale is chosen such as to maximise the number of peaks at ${\rm S/N}\geq3$. We measure the peak distribution of each scale in the observations and in the simulations, again with 5 random realisations of shape noise. The multi-scale cosmological inference is done in the same way as for the single scale but with a data vector which is the concatenation of the data vectors of the individual scales. The data vector contains 60 elements for the combination of the five scales, which is small enough compared to the number of simulations ($N_{\rm s}=175$) to compute accurate constraints with the \citet{Sellentin+16} likelihood.

The joint correlation matrix is shown in Fig.~\ref{fig:corcov5}. This is the mean correlation over five realisations of shape noise, each of which contains the 175 fiducial mocks. We see that the different scales are highly correlated to one another, and also that the closer scales show more correlations. The correlations are larger at large scales, since we increase the scale linearly while the number of galaxies included in the aperture scales with the area, the difference between $\theta=6\arcmin$ and $\theta=9\arcmin$ is therefore larger than the difference between $\theta=15\arcmin$ and $\theta=18\arcmin$.

We compute the constraints on $S_8$ from each individual scale and for different combinations of scales. While all estimates are consistent with the fiducial single scale analysis, the improvement in precision is at best of $\sim10\%$. This value is reached when using the two scales $\theta=12.5\arcmin$ and $\theta=15\arcmin$ together. Adding extra scales does not improve the constraints further, such that it is not necessary to combine more than two scales. This is supported by the large amount of correlation between scales found in the correlation matrix (see Fig.~\ref{fig:corcov5}). The constraints might even get less precise when adding extra scales, probably because of anti-correlations between scales. For example, the combination of the five scales yields marginally better constraints than the fiducial scale alone. In addition, we note that the single-scale analyses yield the most precise constraints for $\theta=12.5\arcmin$ as expected by definition of our fiducial scale. Because multi-scale constraints are only mildly better than the single-scale case, we recommend using only one scale to save computation time. We note that \citet{Liu+15j} also conducted a multi-scale analysis and found that combining more than two scales does not improve the 2D contours in the $(\Omega_{\rm m}, \sigma_8)$ plane further. Their study also seems to show that the multi-scale approach is only marginally better in terms of $S_8$ estimates than the single-scale method.

\subsection{Peak statistics and 2PCFs joint analysis}
\label{sec:discussion:joint}

While peaks represent a different statistic than the 2PCFs, they are both sensitive to LSS and therefore probe correlated information. In this section, we use the 2PCFs as the statistics for cosmic shear and find its joint cosmological constraints with peak statistic. The 2PCFs are defined as

\begin{equation}
\label{eq:2pcf}
\xi_\pm \left( \theta=|\bmath{\theta_{\rm a}}-\bmath{\theta_{\rm b}}| \right)=\frac{\sum_{a,b} w_a w_b \left[ \epsilon_{\rm t}(\bmath{\theta_{\rm a}}) \epsilon_{\rm t}(\bmath{\theta_{\rm b}}) \pm \epsilon_{\times}(\bmath{\theta_{\rm a}}) \epsilon_{\times}(\bmath{\theta_{\rm b}}) \right]}{\sum_{a,b} w_a w_b},
\end{equation}

\noindent where the sum is over pairs of galaxies $a$ and $b$ with separation $|\bmath{\theta_{\rm a}}-\bmath{\theta_{\rm b}}|$ and {\it lens}fit weights $w_a$ and $w_b$. $\epsilon_{\rm t}$ (resp. $\epsilon_{\times}$) represents the ellipticity component tangential (resp. perpendicular) to the line between the two galaxies. Shear two-point correlation functions relate to cosmological parameters through the matter power spectrum \citep[see e.g., ][]{BS01,Kilbinger15}.

\begin{figure}
\centering
\includegraphics[width=0.5\textwidth,clip,angle=0]{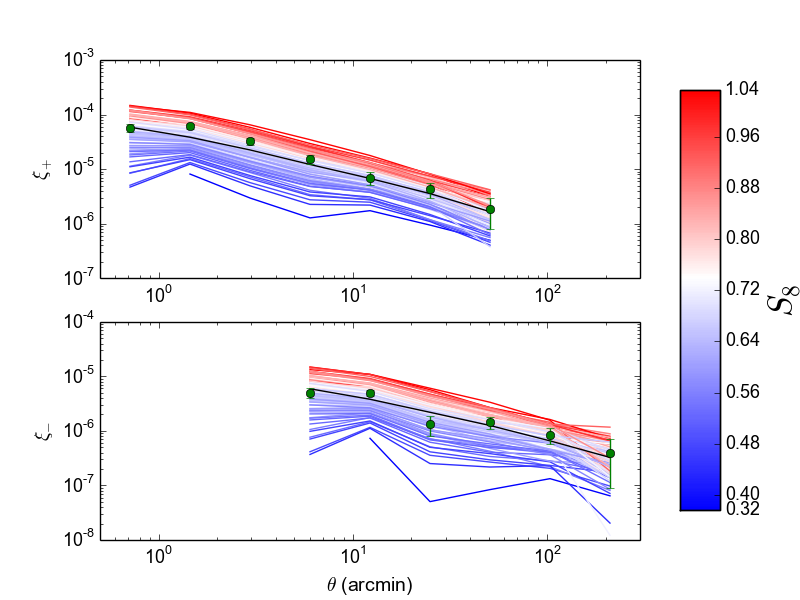}
\caption{Shear two point correlation functions: $\xi_+$ ({\it top}) and $\xi_-$ ({\it bottom}). Green dots represent KiDS data with error bars from the dispersion in the fiducial cosmologies, black line the mean of the fiducial cosmologies, and coloured lines the various simulations with $S_8$ increasing from blue to red.}
\label{fig:distrixi}
\end{figure}

We measure the $\xi_+$ and $\xi_-$ correlation functions using the {\small ATHENA} software \citep{Kilbinger+14} with 900 logarithmic bins in the range $0.5\arcmin \leq \theta \leq 300\arcmin$ and rebin to 9 points logarithmically spaced between the separation limits. In contrast to other KiDS-450 2nd-order cosmic shear studies \citep{Hildebrandt+17,Koehlinger+17,vanUitert+17} which follow a tomographic approach, we use a single redshift bin in the range $0.1 \leq z_{\rm B} \leq 0.9$. The first reason for this is that we want to compare peaks and 2PCFs with similar approaches, and the second is that in this analysis we get the cosmological constraints from comparing the 2PCFs of the observations with that of the simulations, and the simulated mocks we are using are not suited for the tomographic approach. This analysis is also different from the other KiDS cosmic shear papers, as we derive the constraints from N-body simulations and not from an analytical prescription. 
We prefer to use the simulations in this study because we are only interested in the qualitative improvement from the combination of constraints and to ensure that systematics from the simulation approach affect both peaks and 2PCFs measurements. Also, only two cosmological parameters are allowed to vary ($\Omega_{\rm m}$ and $\sigma_8$), rather than 5 or more in the other KiDS-450 cosmic shear studies.

We measure the 2PCFs in the observation and in the \citet{DH10} simulations which follow the KiDS footprint, with same weights and shape noise amplitude as in the data. We note that the 2PCFs do not depend on galaxy positions such that we can use the same positions as in the data in the simulations without biasing the 2PCFs. This allows us to measure the 2PCFs on the exact same mocks as for the peaks, which is important to assess the level of correlation between the two probes.

Following \citet{Hildebrandt+17}, we use only the 7 first bins of $\xi_+$ and the last 6 bins of $\xi_-$. These are displayed in Fig.~\ref{fig:distrixi} as measured in the data (green dots) with error bars from the diagonal elements of the fiducial covariance matrix, in the fiducial simulations (black line for the mean) and in the various cosmologies ranging from low $S_8$ (blue) to high (red). As for the case of peaks (see Fig.~\ref{fig:distripeak}), we find a clear dependence on cosmology, with higher $S_8$ corresponding to higher level of correlation of the shear. The shear correlation in the data is also slightly higher than the fiducial cosmology favouring a moderately higher $S_8$ value. We note that at the largest scale, $\xi_-$ presents large error bars. This is because the simulations we use are only $6\times 6$ square degrees and we are probing correlations between pairs separated by as much as 5 degrees, significantly lowering the number of pairs compared to smaller scales.

\begin{figure}
\centering
\includegraphics[width=0.5\textwidth,clip,angle=0]{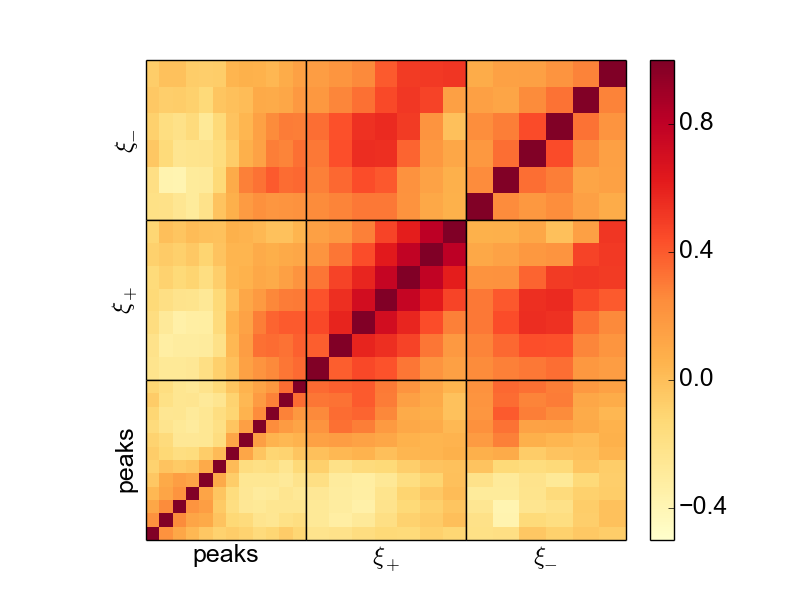}
\caption{Joint correlation matrix for peaks in the range $0\leq {\rm S/N} \leq 4$, $\xi_+$, and $\xi_-$. Each square represents the correlation of one estimator.}
\label{fig:jointcorcov}
\end{figure}

The joint correlation matrix is given in Fig.~\ref{fig:jointcorcov}, for the 175 fiducial mocks, and averaged over five random realisations of shape noise. The peaks span the range $0\leq{\rm S/N}\leq4$, and $\xi_+$ and $\xi_-$ correspond to the 7 and 6 aforementioned bins, respectively. We see a strong correlation between the different scales of $\xi_+$ and also between $\xi_+$ and $\xi_-$. This is also seen in the correlation matrix presented in the appendix of \citet{Hildebrandt+17} for the tomographic case. We also find some correlation between the high-SN peaks and the small-scale $\xi_+$ and $\xi_-$. This is expected as the peaks are probing LSS on scales of about the size of the filter applied in the aperture mass. However, at other scales we note that the correlations remain low, highlighting that the peak statistics are largely independent from the 2PCFs.

We build new data vectors to estimate cosmological constraints with the same method as in the rest of the paper. We use the \citet{Sellentin+16} likelihood with the data, the covariance matrix from the mocks with fiducial cosmologies, and the model from the simulations with various cosmologies interpolated from the mesh at which simulations exist. We probe the constraints from the 2PCFs alone with the concatenation between $\xi_+$ and $\xi_-$ as the data vector, and also the joint constraints with the concatenation of peaks between $0\leq {\rm S/N}\leq 4$, $\xi_+$ and $\xi_-$. The number of bins of these data vectors are respectively 13 and 25, which is still reasonably low compared to the 175 realisations of the fiducial cosmologies used to estimate the covariance matrix.

\begin{figure}
\centering
\includegraphics[width=0.5\textwidth,clip,angle=0]{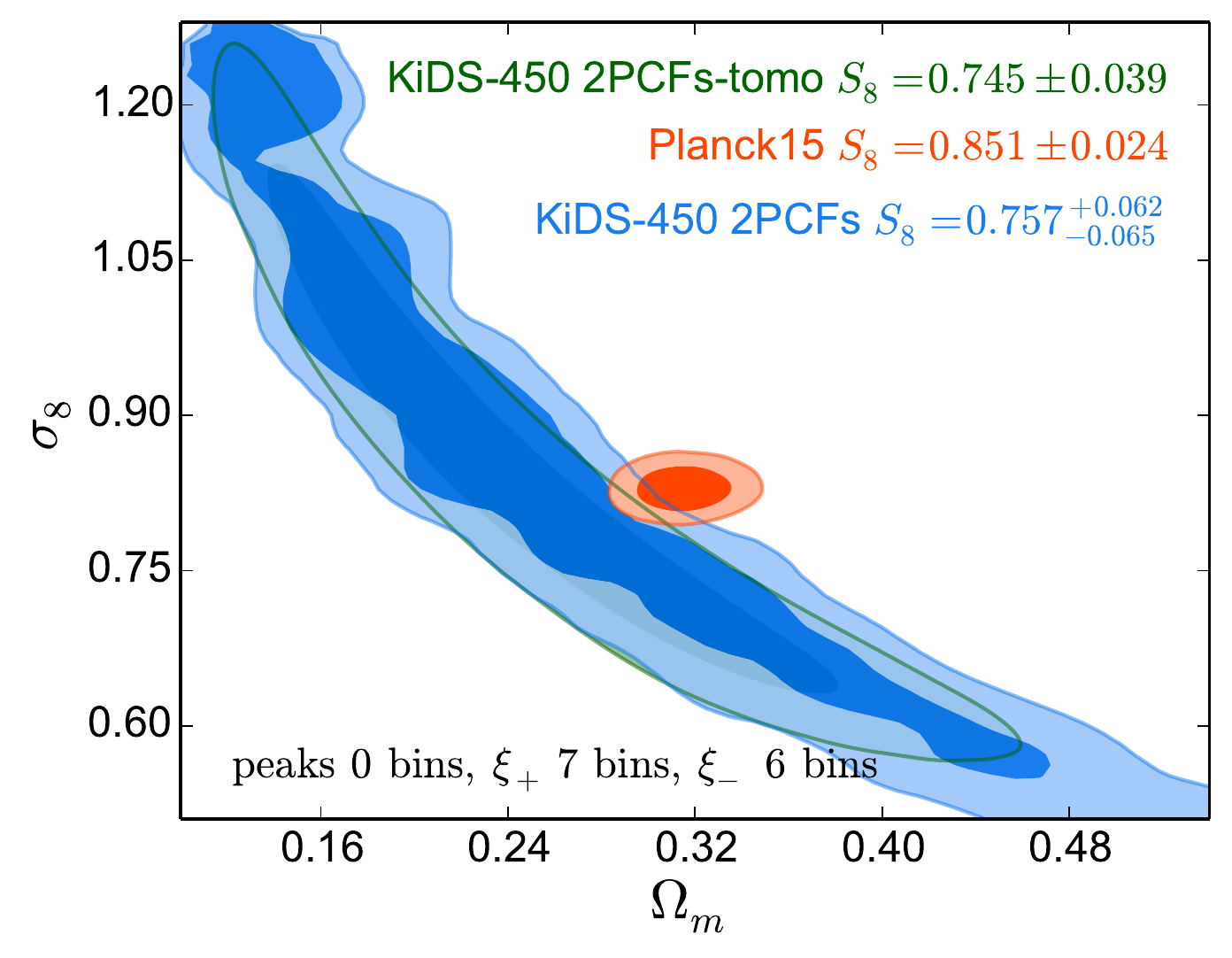}
\includegraphics[width=0.5\textwidth,clip,angle=0]{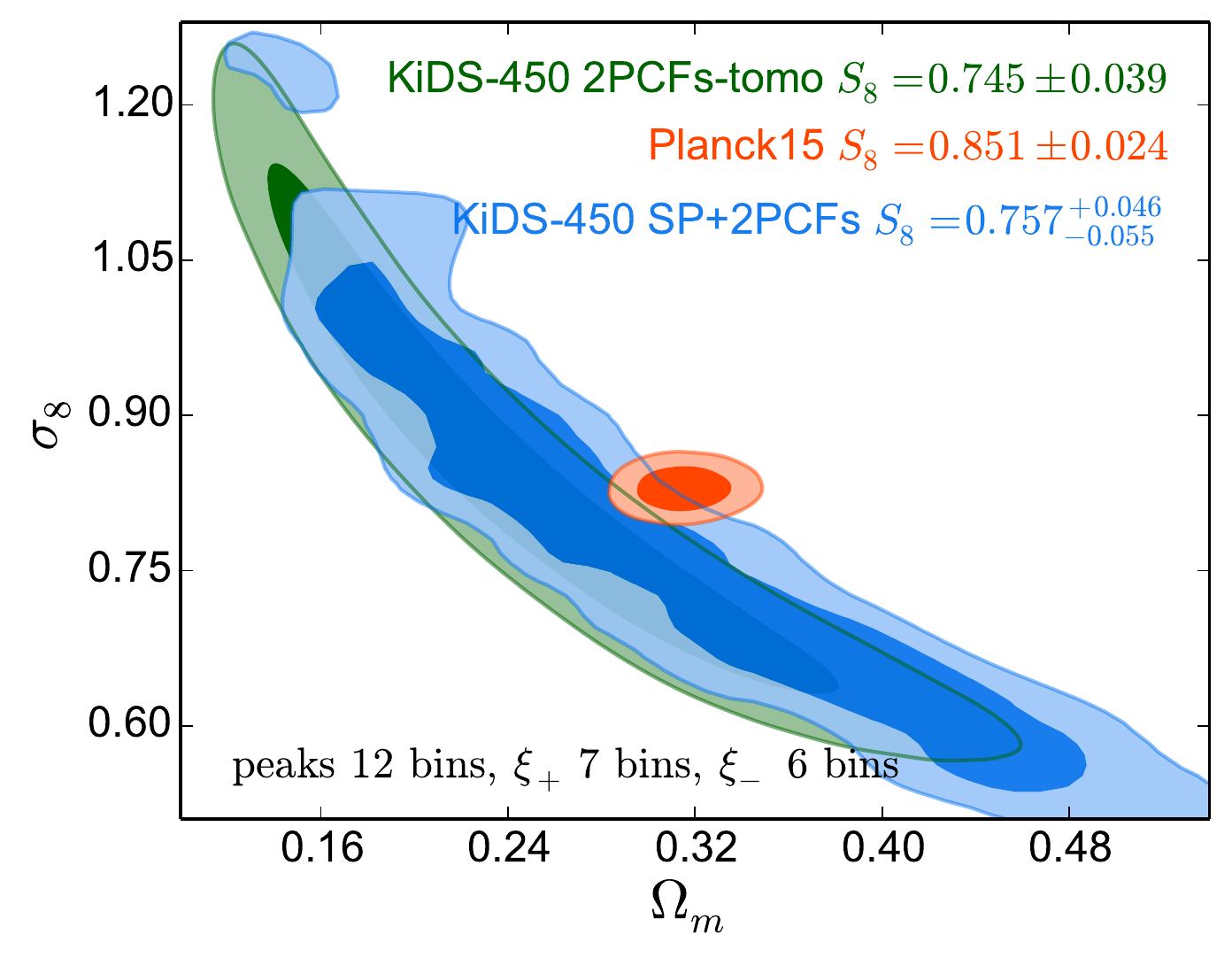}
\caption{Cosmological constraints on ($\Omega_{\rm m}, \sigma_8$) from peaks and 2PCFs. 1$\sigma$ and 2$\sigma$ confidence regions are displayed in dark and light areas respectively. Green contours correspond to KiDS-450 tomographic 2PCFs constraints marginalised over systematics \citep{Hildebrandt+17}, blue to the KiDS-450 2PCFs constraints ({\it top}), and to the joint peaks in the range $0 \leq {\rm S/N} \leq 4$ and 2PCFS constraints ({\it bottom}), and red to Planck "TT+lowP" \citep{PlanckXIII}. The top right legend shows the $S_8$ best value and 68\% errors for each study.}
\label{fig:jointcosmores}
\end{figure}

Figure~\ref{fig:jointcosmores} shows the constraints for the 2PCFs and the joint constraints. There is a very good agreement between the present non-tomographic 2PCFs constraints and the tomographic constraints of \citet{Hildebrandt+17} both in the ($\Omega_{\rm m}, \sigma_8$) degeneracy and in the $S_8$ estimate. Our errors are however $\sim39\%$ larger because we do not use the information from the different redshifts. Quantitatively, the constraints from 2PCFs only are $S_8=0.757^{+0.062}_{-0.065}$, which is the same value as for the peaks but with $\sim16\%$ larger statistical errors. This highlights the very high potential of peak statistics as a cosmological probe compared to the classical WL probe. Furthermore, the combination of both yields an $\sim20\%$ improvement compared to the 2PCFs alone with $S_8=0.757^{+0.046}_{-0.055}$ but no significant improvement from peaks alone. We however stress that this study presents some limitations. First, the likelihood is quite noisy due to sparsity in the probed cosmologies. Second, no systematics are accounted for in this part of the discussion, so only the statistical errors are considered. This is because the impact of systematics on constraints from peak statistics is not known with the same accuracy as that of the 2PCFs, mainly because of the very recent development of peak statistics.

Our results are nonetheless very promising for peak statistics and call attention to the great interest of developing peaks further in terms of systematics comprehension. Our study also confirm the predictions from simulation-based analyses of the improvement of constraints from joint 2nd-order and higher-order cosmic shear over 2nd-order alone \citep[e.g., ][]{DH10,Hilbert+12}. Finally, we note that in their study of CFHTLenS, \cite{Liu+15j} also found only marginal improvement from adding convergence peaks to the convergence power spectrum compared to peaks alone, and $\sim 40\%$ improvement compared to power spectrum alone, taking the full covariance between the two probes into account as we do here.

\section{CONCLUSION}
\label{sec:conclusion}

In this paper, we derive constraints on $S_8=\sigma_8 \sqrt{\Omega_{\rm m}/0.3}$ by comparing the peak statistics in the WL mass map of KiDS-450 to that of simulations with various cosmologies. Mass maps are computed by averaging the tangential shear in a 12.5 arcmin radius aperture with an NFW-like weighting function, which is compensated in the convergence field to avoid the mass sheet degeneracy.

We find constraints on $S_8$ in good agreement with those from 2PCFs \citep{Hildebrandt+17}, and also from the independent peak statistics study of \citetalias{Shan+17}. The latter uses convergence peaks and analytical predictions, focusing on the high-S/N peaks corresponding to high-mass halos. Our $S_8$ estimate is 1.6$\sigma$ lower than the value estimated with Planck CMB when we account for systematics. We also measure the gain of information when adding the low-S/N peaks, which correspond to projections of low-mass halos, to the high-S/N peaks, corresponding to high-mass halos. Quantitatively, the $S_8$ estimate improves by $\sim25$\% when adding peaks with S/N lower than 3. We measure the peak distribution with various filter scales finding only a mild ($\sim 10\%$) improvement from combining scales. Refining the covariance matrix to properly account for sample variance only affects cosmological constraints at the level of a few percents, validating the fiducial approach of this paper. Finally, we measure the non-tomographic 2PCFs and find consistent $S_8$ estimates between peaks and 2PCFs. Combining both probes yields an $\sim 20$\% improvement compared to 2PCFs alone, highlighting the high potential of peak statistics for future WL surveys.

\section*{Acknowledgements}

We thank Tomasz Kacprzak, Tim Schrabback, Patrick Simon, and Angus Wright for useful discussions. We additionally thank the referee for interesting comments and suggestions. HHi acknowledges support from the DFG under Emmy Noether grant Hi 1495/2-1. MA and CH acknowledge support from the European Research Council under grant number 647112. HHo acknowledges support from Vici grant 639.043.512, financed by the Netherlands Organization for Scientific Research (NWO). This work is supported by the Deutsche Forschungsgemeinschaft in the framework of the TR33 ``The Dark Universe'' (PS and DK). The research leading to these results has received funding from the European Union's FP7 and Horizon 2020 research and innovation programs under Marie Sklodowska-Curie grant agreement numbers 627288 and 664931 (JM). JHD acknowledges support from the European Commission under a Marie-Sk{l}odwoska-Curie European Fellowship (EU project 656869). KK acknowledges support by the Alexander von Humboldt Foundation. RN acknowledges support from the German Federal Ministry for Economic Affairs and Energy (BMWi) provided via DLR under project no. 50QE1103. Computations for the SLICS $N$-body simulations were performed in part on the Orcinus supercomputer at the WestGrid HPC consortium (\url{www.westgrid.ca}), in part  on the GPC supercomputer at the SciNet HPC Consortium. SciNet is funded by: the Canada Foundation for Innovation under the auspices of Compute Canada; the Government of Ontario; Ontario Research Fund - Research Excellence; and the University of Toronto.




\bibliographystyle{mnras}
\bibliography{SuPerKiDS}




\appendix

\section{Evaluating the interpolation of the peak distribution}
\label{app:interp}

\begin{figure}
\centering
\includegraphics[width=0.7\textwidth,clip,angle=0]{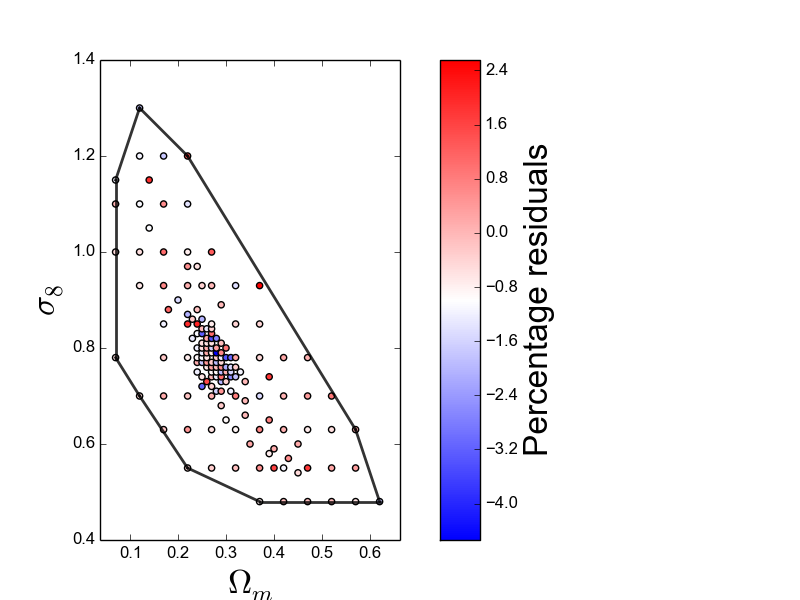}
\caption{Percentage residuals from the interpolation of the number of peaks in the bin $3.33 \leq {\rm S/N} \leq 3.66$. The black polygon represents the convex hull within which we trust the interpolation.}
\label{fig:interp2}
\end{figure}

\begin{figure}
\centering
\includegraphics[width=0.5\textwidth,clip,angle=0]{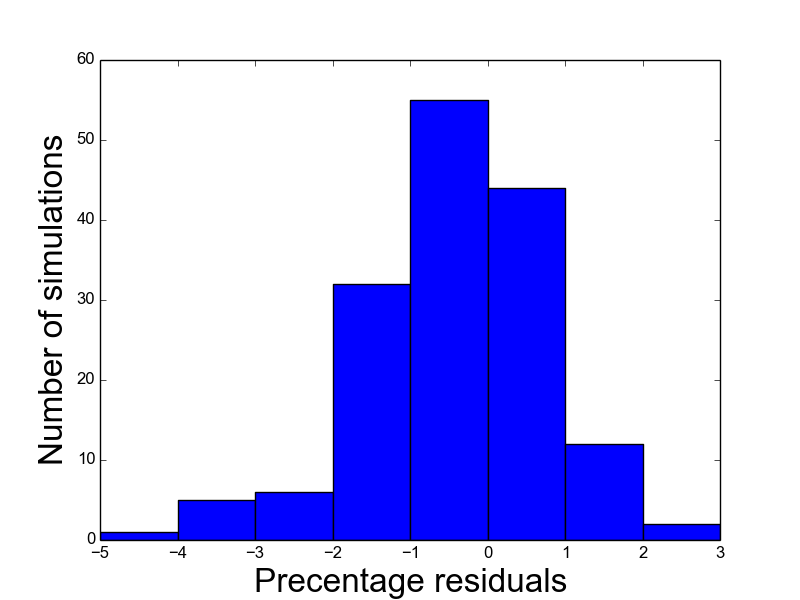}
\caption{Distribution of the percentage residuals from the interpolation of the number of peaks in the bin $3.33 \leq {\rm S/N} \leq 3.66$.}
\label{fig:interp3}
\end{figure}

In Sect.~\ref{subsec:interp} we describe how we interpolate the peak distribution from our 157 different points in the ($\Omega_{\rm m}, \sigma_8$) plane to an evenly spaced grid with a step of 0.01 for both parameters. To evaluate the robustness of this interpolation we compare the interpolated numbers of peaks with the measured ones for every simulation. The relative difference is displayed in Fig.~\ref{fig:interp2} for the same bin of S/N as that of Fig.~\ref{fig:peaksint}: $3.33 \leq {\rm S/N} \leq 3.66$. The percentage residual is always lower than 5\% and ranges from $-5$ to $+3$\%. The number of peaks in this bin being roughly 450, the Poisson error is about 5\%. The error in the interpolation process is therefore always lower than the  Poisson error. In addition we calculate the distribution of the residuals (Fig.~\ref{fig:interp3}) and find that it is centred around 0\% and that more than 85\% of the simulations have a relative error due to the interpolation of less than 2\%. This shows that the interpolation process does not add any significant systematic bias.

\section{Negative peaks}
\label{app:negpeaks}

\begin{figure}
\centering
\includegraphics[width=0.5\textwidth,clip,angle=0]{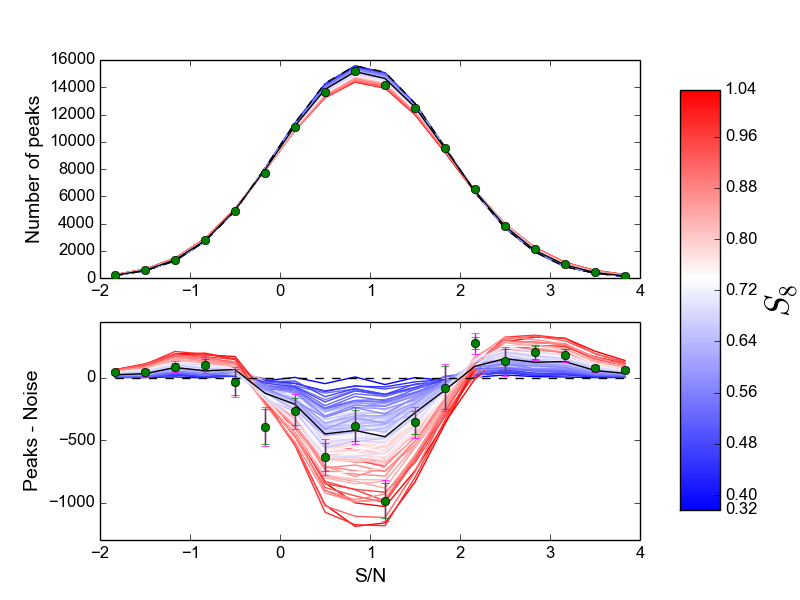}
\caption{Same as Fig.~\ref{fig:distripeak} but for peaks in the range $-2\leq {\rm S/N} \leq 4$.}
\label{fig:distri_yi}
\end{figure}

\begin{figure}
\centering
\includegraphics[width=0.5\textwidth,clip,angle=0]{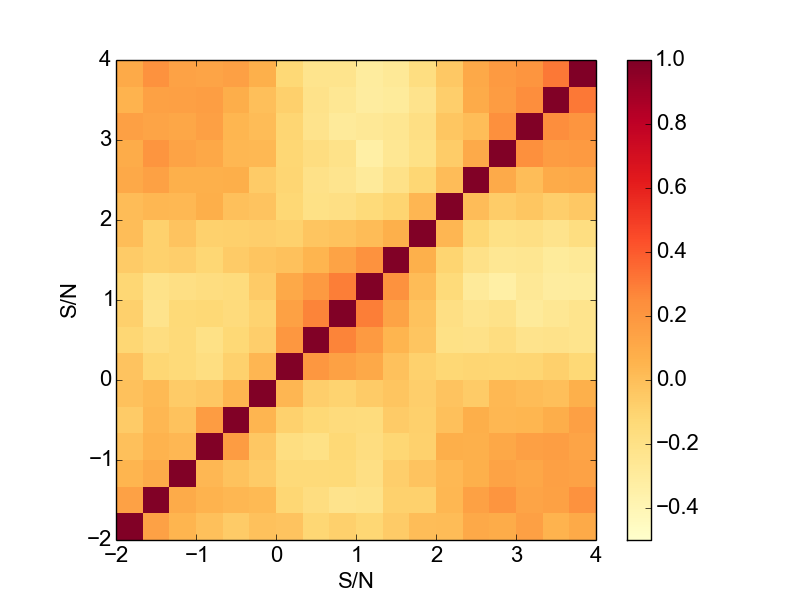}
\caption{Correlation matrix for peaks in the range $-2\leq {\rm S/N} \leq 4$.}
\label{fig:corcov_yi}
\end{figure}

In this appendix we study the possible gain of information from using the negative S/N peaks. We build the data vector with adding all negative bins that pass the Gaussian variable assumption. This gives us a new vector with 20 bins in the range $-2\leq {\rm S/N} \leq 4$, shown in Fig.~\ref{fig:distri_yi}. We note that the excess peak distribution over the noise peaks in the negative bins presents a similar shape as that of the positive peaks but with about twice smaller values and extending to less negative values. We then compute the constraints on $S_8$ using the new peak distribution and obtain $S_8=0.767^{+0.050}_{-0.065}$ without taking systematics into account. These constraints are of the same order as those from the positive peak distribution only ($S_8=0.757^{+0.054}_{-0.053}$). This can be understood by looking at the correlation matrix from the full peak distribution in Fig.~\ref{fig:corcov_yi}. We see that the negative peaks with $-2\leq {\rm S/N} \leq 0$ are highly correlated with those of $2\leq {\rm S/N} \leq 4$, explaining why we do not gain information by including them.

\section{Refined covariance matrix}
\label{sec:discussion:cov}

In this section, we use the set of mocks described in Sect.~\ref{subsec:slics} to measure the impact of simulation types on the derived cosmological constraints, and validate part of the analysing process. The SLICS mocks benefit from several improvements compared to those of \citet{DH10} used in the rest of the paper. First, the SLICS simulations \citep{Harnois-Deraps+15} have a higher resolution with $1536^3$ particles against $256^3$, extend to $z=3$, and cover 100~deg$^2$ instead of 36~deg$^2$. In addition, having access to the full shear planes, we could tailor the simulations specifically for our project and populate the maps with galaxies following the position and $n(z)$ of the KiDS-450 data, without having to rely on the interpolation scheme that we applied to the \citet{DH10} mocks. We also extend the redshift range to the full DIR redshift distribution of KiDS-450 while the \citet{DH10} mocks have almost no galaxies at $z>2$ due to the redshift distribution they used, although the simulations also extend to $z=3$. Sample variance is also better included in these refined mocks by using different N-body simulations to tile the KiDS-450 area, instead of repeating a single simulation across this area.

We run the same algorithm to identify peaks in these mocks, and derive the cosmological constraints using the covariance matrix from this set of simulations but still using the \citet{DH10} simulations to compute the model of the peak dependence on cosmology. We use 5 random realisations of shape noise, the same number as in our fiducial analysis. The number of fiducial mocks is 67 which is lower than the 175 of the main analysis but each of these mocks now better accounts for sample variance. We also note that the cosmological parameters are slightly different in the fiducial SLICS than in the \citet{DH10} simulations, but we do not expect a large variation of the covariance matrix with cosmology.

\begin{figure}
\centering
\includegraphics[width=0.5\textwidth,clip,angle=0]{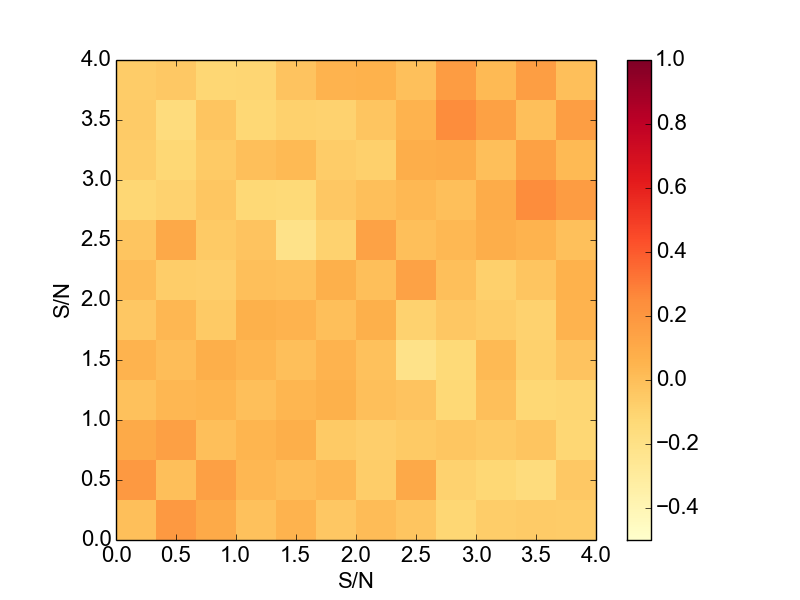}
\caption{Difference between the fiducial correlation matrix of Fig.~\ref{fig:corcov} and the refined correlation matrix for peaks in the range $0\leq {\rm S/N} \leq 4$.}
\label{fig:corcov_sm}
\end{figure}

In Fig.~\ref{fig:corcov_sm}, we display the difference between the correlation matrix of the first set of simulations shown in Fig.~\ref{fig:corcov} and the refined correlation matrix. The agreement between the two correlation matrices is good, both presenting low correlation between peaks, with somewhat higher correlations at the high S/N-peaks, leading to homogeneous residual correlation. However, the new correlation matrix shows higher scatter than the previous one, and lower correlations, leading to residual correlation of up to 0.2. This can be attributed to the larger area covered by the SLICS simulations and the proper handling of sample variance, which provides us with a more representative population of peaks. We also re-computed the \citet{DH10} correlation matrix with only 67 simulations finding that the observed differences are not due to the use of different number of simulations.

As described in Sect.~\ref{subsec:res}, the constraints on $S_8$ using the refined covariance matrix are almost identical to that of the main analysis. This validates the various approximations we made when building the mocks from the \citet{DH10} simulations.


\bsp	
\label{lastpage}
\end{document}